\documentclass[preprint2]{aastex}

\usepackage{natbib}
\usepackage{graphicx}
\usepackage{epstopdf}

\newcommand{\epeak}{$E_{\rm peak}$}

\shorttitle{Which \epeak?}
\shortauthors{Preece, Goldstein et al.}

\begin{document}

%%Title
\title{Which \epeak? -- The Characteristic Energy of Gamma-Ray Burst Spectra}

%%Authors
\author{Robert Preece}
\affil{Space Science Department, The University of Alabama in Huntsville,
    Huntsville, AL 35809}

\author{Adam Goldstein}
\affil{Space Science Office, VP62, NASA/Marshall Space Flight Center, Huntsville, AL 35812}

\author{Narayana Bhat and Matthew Stanbro}
\affil{Center for Space Plasma and Aeronomic Research, The University of Alabama in Huntsville,
    Huntsville, AL 35809}
    
\author{Jon Hakkila}
\affil{The University of Charleston, SC at the College of Charleston 29424}

\and

\author{Dylan Blalock}
\affil{Physics Department, The University of Alabama in Huntsville,
    Huntsville, AL 35809}

%%Abstract
\begin{abstract}
A characteristic energy of individual gamma-ray burst (GRB) spectra can in most cases be determined from the 
peak energy of the energy density spectra ($\nu{\cal F}_{\nu}$), called `\epeak'. Distributions of 
\epeak\ have been compiled for time-resolved spectra from bright GRBs, and also time-averaged spectra and peak 
flux spectra for nearly every burst observed by CGRO-BATSE and Fermi-GBM. Even when determined by an 
instrument with a broad energy band, such as GBM (8 keV to 40 MeV), the distributions 
themselves peak at around 240 keV in the observer's frame, with a spread of roughly a decade in energy. 
\epeak\ can have considerable evolution (sometimes greater than one decade) within any given burst,  
as amply demonstrated by single pulses in GRB110721A and GRB130427A. Meanwhile, several luminosity 
or energy relations have been proposed to correlate with either the time-integrated or peak flux \epeak. 
Thus, when discussing correlations with \epeak, the question arises, ``Which \epeak?''. A single burst may be 
characterized by any one of a number of values for \epeak\ that are associated with it. Using a single pulse 
simulation model with spectral evolution as a proxy for the type of spectral evolution observed in 
many bursts, we investigate how the time-averaged \epeak\ emerges from the spectral evolution within a 
single pulse, how this average naturally correlates with the peak flux derived \epeak\ in a burst and 
how the distribution in \epeak\ values from many bursts derives its surprisingly narrow width. 
\end{abstract}


%%Introduction
\section{Introduction}

%There seems to be good evidence that the observed distributions in the values for \epeak, whether 
%that of time-integrated, peak flux or time-resolved spectra, are intrinsically narrow 
%\citep{GoldsteinCat,GruberCat,KanekoCat,NavaCat}, with a full width at half 
%maximum (FWHM) of roughly a decade in energy from instruments with energy coverage that spans two decades. 
%This result has largely held firm in the published results from the Gamma-ray Burst Monitor (GBM) on board 
%the Fermi Gamma-ray Space Telescope mission.

Unquestionably, a characteristic energy can be an important diagnostic feature of the spectra of 
cosmological sources. If the sources are standard in some way, with identifiable atomic or nuclear lines, 
for example, the energy measurement could then directly correspond 
with distance via the cosmological redshift. From early on, analyses of the prompt emission  
of gamma-ray bursts (GRBs) revealed that a spectral break typically occurred within the instrumental 
pass band of scintillator-based detectors -- 
between approximately 20 keV to 1 MeV \citep{Mazets,Metzger,Harris}. 
The spectral break can take on the form of an exponentially-attenuated power-law; in which case, it 
had been associated with an optically-thin thermal bremsstrahlung temperature.  
However, many burst spectra are typically harder than thermal, with a 
high-energy tail that may be characterized by a power law. This led \citet{band93} to propose an 
entirely phenomenological spectral form that is the unique function joining two power laws 
that is continuous and smooth to first order: the `Band' GRB function. This function finds its most 
familiar form in equation \ref{bandmodel}, in a 
parametrization where the break energy is cast as the peak in the spectral energy distribution 
($E^2 f_{\rm Band}(E)$ or $\nu {\cal F}_{\nu}$): \epeak.
\begin{eqnarray}
f_{\rm Band}(E) & = & A \left( \frac{E}{100 \; {\rm keV}}\right)^{\alpha} 
\exp \left(-\frac{E\; (2+\alpha)}{E_{\rm peak}}\right)\nonumber\\
{\rm if} \quad E & < & \frac{(\alpha-\beta)\; E_{\rm peak}}{(2+\alpha)} 
\equiv E_{\rm break} {\rm ,}\nonumber\\
\quad f_{\rm Band}(E) & = & A \left[\frac{(\alpha-\beta) \; E_{\rm peak}}
{(2+\alpha)\; 100\; {\rm keV}}\right]^{(\alpha-\beta)}\label{epmod} \times \nonumber\\
&  & \exp{(\beta-\alpha)} \left(\frac{E}{100 \; {\rm keV}}\right)^{\beta}\nonumber\\
{\rm if} \quad E & \geq & \frac{(\alpha-\beta)\; E_{\rm peak}}{(2+\alpha)}{\rm .} \label{bandmodel}
\end{eqnarray}
From the form of the function, both $\alpha$ and $\beta$ are power law indices in photon number. 
Because of the pre-factor $E^2$ in the definition of $\nu {\cal F}_{\nu}$, \epeak\ is only defined for 
values of $\beta$ less than $-2$ (as is typical). For large negative values of $\beta$, 
we recover thermal-like spectra, guaranteed to yield a value for \epeak\ as long as $\alpha > -2$. 
With three shape parameters, the Band function is flexible enough to stand in for a number of physical 
processes likely to occur in GRBs, including blackbody emission, synchrotron emission by thermal or 
power law electrons, as well as Compton scattering in any of its various guises. However, the Band function has a specific 
constant curvature that does not match any of these in detail, especially in the region where the two 
power law segments join (e.g.: see figure 8 in \citet{burgess14}).

For the purposes of the current work, it is more 
convenient to use a cut-off model power-law (the so-called COMP function in the spectral fitting 
package RMFIT\footnote{RMFIT version 4.4.2 was used throughout this work. RMFIT is publicly available 
at NASA's HEASARC website: http://Fermi.gsfc.nasa.gov/ssc/data/analysis/rmfit.}), which is identical to 
the Band function in the limit $\beta \rightarrow -\infty$. The COMP functional 
form can also be cast in terms of \epeak, as shown in equation \ref{comp}.
\begin{equation}
f_{\rm COMP}(E) = A \left( \frac{E}{100 \; {\rm keV}}\right)^{\alpha} 
\exp \left(-\frac{E\; {(2+\alpha)}}{E_{\rm peak}}\right){\rm .} \label{comp} %\nonumber\\ E_0 {(2+\alpha)} & \equiv &  E_{\rm peak}
\end{equation}
There are several reasons that drive our choice of this function for the simulations described below. First, for most 
bursts where the spectral evolution can be determined as a function of time, it is possible to create 
time slices that can track the evolution of \epeak\ (and $\alpha$) with good precision, while for the 
same temporal accumulations, $\beta$ can not be determined with any precision. The reason for this 
is that both the effective area and the flux fall rapidly with increasing energy, resulting in too 
few counts above \epeak\ to determine the spectrum. Indeed, one of the results from the Compton Observatory's 
Burst And Transient Source Experiment (BATSE) 
Catalog of time-resolved spectroscopy of bright bursts \citep{KanekoCat} was that the COMP function was 
preferred over the Band function in a majority of the spectra fit (34.9\% vs.\ 33.7\% for the BEST model; 
see Table 8 in the Catalog). It is possible that some GRB spectra arise from synchrotron emission from 
thermal electrons, especially just after the photosphere has become optically thin. In that case, the 
emission should resemble the COMP function with $\alpha \sim -2/3$.
Even when a Band function fit is preferred, for example, as in the spectral 
analysis of the first pulse of the bright GRB 130427A \citep{preece2014}, the Band $\beta$ parameter was 
required to be held fixed for each 0.1 s accumulation; otherwise, the extra free parameter creates instabilities 
in the fits to the other parameters, due to their high degree of cross-correlation (as noted in, e.g., \citet{KanekoCat}). 
Finally, the sum of several exponentials evolving under under certain conditions can approximate a power law, as we 
shall show below. Conversely, the sum of many power laws with different spectral indicies is guaranteed not to result in a power law.
%; generally, the sum results in a concave upward curve, not unlike what has been observed in several bursts 
%\citep{Gonzalez,GRB090510,GRB090926A}\footnote{It should be noted that the change in power law slope from $< -2$ 
%to $\sim - 1$ with increasing energy observed in these bursts is much easier explained as the sum of 
%two components than the average of many spectra evolving from one power index to the other.}. 
%As the Band function is composed of two power law segments, it poses a doubly special 
%difficulty for spectral evolution (but see GRB 090902B: Figure 3 of \citet{GRB090902B}). 
It is one of 
the outstanding questions of GRB spectroscopy why the data should ever be consistent with the Band function.

As mentioned above, assuming it represents a characteristic energy, \epeak\
is the unique spectral parameter that encodes relative motion between the source and the observer. 
Given that GRB redshifts have been measured for many events, the distribution of these currently 
falls between $z \approx 0.01$ \citep{galama98} and $z \approx 9.4 $ \citep{Cucchiara},
with a peak between $z = 1$ and 2. 
% The observed energy should behave as $E_{\rm obs} = E_{\rm src} / (z + 1)$, 
% where the characteristic energy at the source is $E_{\rm src}$. 
Without any 
other consideration, we might expect the distribution of \epeak\ to follow the redshift distribution, 
with the smallest values corresponding with the largest redshifts. 
However, there is good reason to suspect that the GRB emission arises within a relativistic blast 
wave \citep{reesmesz}. Any characteristic energy, as seen in the observer's frame, should thus also 
be multiplied by the bulk Lorentz factor $\Gamma$: 
\begin{equation}
E_{\rm obs} = E_{\rm src} \Gamma / (z + 1), \label{e_obs}
\end{equation}
where the characteristic energy at the source is $E_{\rm src}$.
Under the most common assumptions, a limit on the bulk Lorentz factor can be determined as follows: 
the blast wave must be optically thin to its own photons, otherwise it would still be a fireball, due to 
photon-photon pair production. The emission from a relativistic blast wave is highly beamed, raising 
the pair-production energy threshold to 1 MeV $\Gamma / \gamma > 1$, for photons with energies of 
$\gamma$, in units of the electron's rest energy. A more careful treatment would have to include details 
of the spectrum against which the highest energy photon is scattering 
\citep{BaringHarding97,2006ApJ...650.1004B}. Thus, in the simplest case, if a photon 
with energy $\gamma$ is observed, it puts a limit on the bulk Lorentz factor $\Gamma > \gamma$. 
Thus, $\Gamma$, as determined by the maximum observed photon energy, is found to vary from 
burst to burst. The intrinsic variability of the central engine % In addition, physical processes such as 
or the possibility that relativistic shock collisions accelerate 
particles at the expense of $\Gamma$ \citep{reesmesz}, make it very likely that $\Gamma$ 
is expected to vary {\em during} a burst. All of 
these factors are reflected in the observed values and evolution of \epeak.
%Finally, there is 
%the possibility of evolution of intrinsic \epeak\ with redshift, due to some unknown character of the 
%progenitor, with the result that, all other factors being constant, the average 
%value of \epeak\ over several bursts could change with redshift.

\begin{figure}[h]
\includegraphics[scale=0.5]{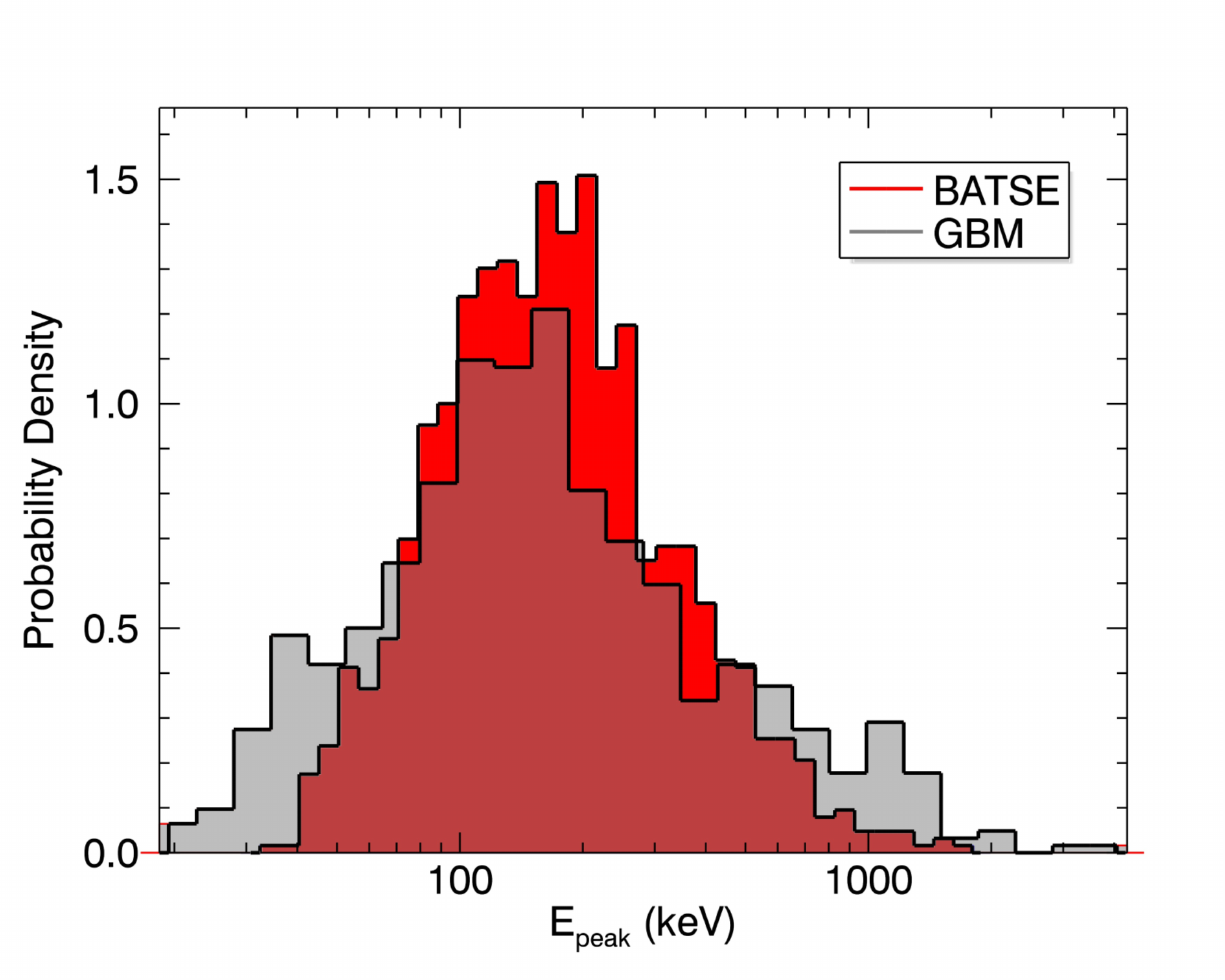}
\caption{Histograms of the \epeak\ values derived from fluence-averaged spectra, comprised of 
1297 BATSE bursts (grey - \citet{GoldsteinCat}) and 680 GBM bursts (red - \citet{GruberCat}). 
The extended spectral coverage of GBM clearly shows up in the enhancement of bursts found in the 
low- and high-energy wings. The widths of the two distributions are qualitatively similar. \label{BATSE_GBM_Epeak}}
\end{figure}

However, it is clearly true that \epeak\ arises from some {\em physical} emission process, so on 
top of the known sources of potential variation on the observed quantity, there is an unknown 
intrinsic distribution in \epeak\ that is related to the underlying physical processes. This 
intrinsic distribution may be quite wide, as can be seen in the time-resolved spectra of single-pulse 
bursts where the redshift is known, like GRB110721A \citep{axelsson} 
and the first pulse of GRB130427A \citep{preece2014}. In each of these, hard-to-soft spectral evolution 
was observed, with \epeak\ varying over considerably greater than one decade in energy. Despite the supposition 
that each of the contributions to the variation in \epeak\ should operate completely independently, the overall 
distribution of fitted values of \epeak\ is less than a decade in width (e.g.: Figure \ref{BATSE_GBM_Epeak}, 
discussed below), whether the distribution 
is drawn from the average spectra of a large sample of bursts \citep{GoldsteinCat,NavaCat,GruberCat} or 
from statistically independent time-resolved spectra from bright bursts \citep{KanekoCat}. 
Much has been made of broad correlations between \epeak\ and other measured quantities, 
such as the total (isotropic) energy, as in the 
relation of \citet{Amati02}. Such relations could be useful for obtaining red-shifts directly 
from spectral measurements, for example. However, several groups have analyzed such relations and  
claim that they are artifacts of selection effects \citep{NakarPiran,BandPreece,SchaeferCollazzi}. 

For those spectra where \epeak\ values could be determined reliably, the distribution 
roughly resembles a log-normal function of approximately a decade in width (e.g.: the 'GOOD' set as seen in Figure 11 
from \citet{KanekoCat}). The specific shape of the distribution is not specified by any theory; so the 
paucity of low- and high-energy \epeak\ values has widely been attributed to detector selection effects. 
Still, it is not known why \epeak\ should not be evenly distributed between the lowest and highest 
energy bounds of the detector. It was speculated that once the Fermi Gamma-ray Burst Monitor (GBM) 
flew, the question of the reality of the high-energy cutoff of this distribution would 
be settled \citep{PiranReviewInKouveliotou}. It has; the \epeak\ distribution cutoff for GBM, relative to BATSE, does not 
change all that much, even though GBM has the medium-energy (200 keV to 40 MeV) range of the BGO detectors to contribute to the 
spectral fitting. Figure \ref{BATSE_GBM_Epeak} shows a comparison between the fluence (average spectrum) 
\epeak\ values between the BATSE and GBM data sets \citep{GoldsteinCat,GruberCat}, indicating that GBM extends the distribution by 
including more lower and higher energy values. %, while at the same time generally preserving the width of the distribution. 
That is not to say that there are {\em no} spectra with large values of \epeak, just that 
the majority of fitted \epeak\ values cluster around a ~200 keV value. Certainly, there is no `undiscovered country' of 
\epeak values clustering in the MeV range. We will explore why this might be in this paper.

In order to facilitate understanding of the important role \epeak\ plays in the various observed correlations 
with other burst observables, we examine herein various factors that can affect the observed values 
of \epeak. First of all, in section 2, we examine the limits of detectability of a single value for 
\epeak\ from simulations of spectra of increasingly diminished intensity. Next, in section 3, we 
look at simulations of hard-to-soft pulses and the role of spectral evolution in the determination 
of average values of \epeak. In section 4, we look at simulations of multiple, overlapping hard-to-soft 
pulses. Finally, we examine some intrinsic correlations that arise naturally in pulses with spectral 
evolution.

\section{Limits on \epeak\ Detectability}

\begin{figure}[t!]
\includegraphics[scale=.35]{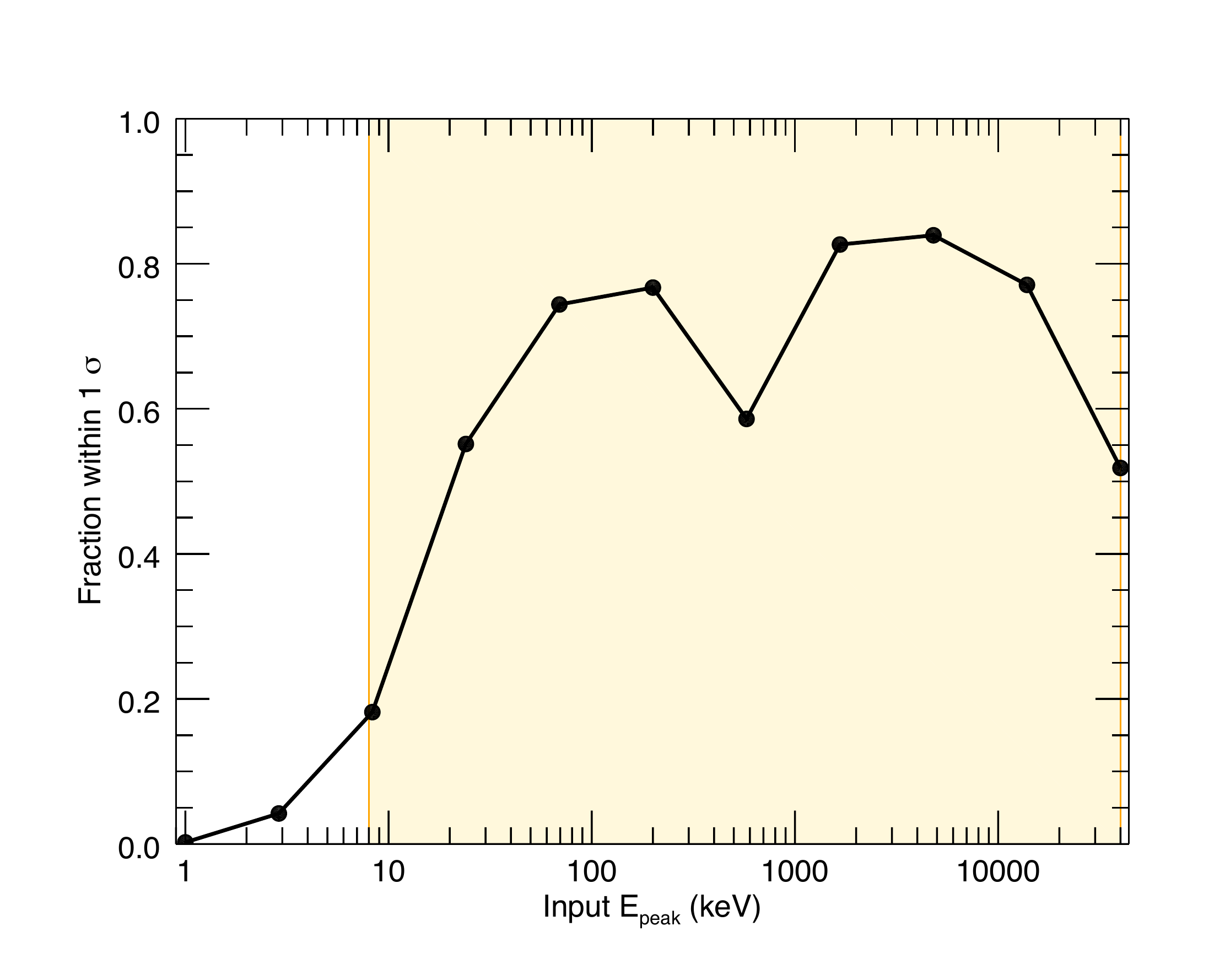}
\caption{Plot of the fraction of Band function fits that contain the simulated input $E_{\rm peak}$ 
energy within the 1$\sigma$ confidence interval of the fits.  The simulations were
performed over a grid of amplitude, alpha, beta, and $E_{\rm peak}$ values and the plot
marginalizes over all parameters except $E_{\rm peak}$.  The orange region denotes the GBM
energy range. \label{EpeakRecover}}
\end{figure}

By simulating $\sim15000$ different spectra via varying all Band function parameters on a
spectral grid using GBM responses, we can study the efficiency of recovering $E_{\rm peak}$. 
Figure \ref{EpeakRecover} shows the fraction of Band function fits that include the simulated true value of
$E_{\rm peak}$ within the 1$\sigma$ uncertainty, marginalizing over all inputs of the
amplitude, alpha, and beta. We find that $E_{\rm peak}$ can be reliably recovered between $\sim
20$ keV--40 MeV.  Considering that the GBM bandpass is 8 keV--40 MeV, nearly all $E_{\rm peak}$
values that exist within the GBM band can be reliably recovered via spectral fitting. 
The low point at $~600$ keV represents a roughly 1 -- 2 $\sigma$ fluctuation.
Because the estimation and constraint of the Band function curvature is dependent on the low-energy
index, $E_{\rm peak}$ values $< 20$ keV observed by GBM are unlikely to be constrained because
there is insufficient data to constrain the low-energy index.  In this case, the spectrum would
appear similar to a simple power law with an index similar to the index of the Band high-energy
index. Figure 27$b$ in \citet{GoldsteinCat} presents a similar set of simulations with BATSE
with similar findings. Due to the narrower energy band of BATSE (20-2000 keV) with respect to GBM, there was also
an apparent decrease in accuracy if $E_{\rm peak}$ existed above the detector band, although
the loss of accuracy was not as dramatic compared to the low-$E_{\rm peak}$ case.  The
interpretation for the high-$E_{\rm peak}$ case is that as long as the low-energy index is
reasonably constrained and some curvature is measurable within the detector band, then it is
possible to estimate $E_{\rm peak}$, albeit with less accuracy and constraint. It is in 
part to avoid the issue of the interplay between fitting \epeak\ and $\beta$ in the Band function 
that we choose to perform our pulse spectral evolution simulations with the COMP function, as discussed above.

\section{Hard-to-soft \epeak\ Evolution in Pulses}

Recent work has firmly established that pulses, defined as an emission episode with a rise in brightness
to a peak followed by a decay, are the prime building blocks that make up the prompt 
emission light curve in GRBs \citep{HakkilaCumbee}. One piece of evidence supporting this conclusion is the 
observation of temporal evolution of \epeak\ throughout many bright, isolated pulses. In the hard-to-soft (HTS) case, \epeak\ evolves 
smoothly (in many cases as a power law) from an initial high value at the onset of the pulse, through 
the rising portion, the peak and the decaying portion to a final low energy when the pulse can no 
longer be detected \citep{lu10,NorrisCat,axelsson,preece2014,burgess14}. 
Another behavior, called hardness-intensity `tracking' (HIT), is where the fitted value of \epeak\ follows 
the pulse intensity (expressed as integrated counts over a standard energy range) with good correlation. 
\citet{HakkilaPreece} and \citet{lu10} have both suggested that the tracking behavior may arise when several pulses 
are run together, so that they many not be clearly distinguished. We will also address this below.
In many cases, the rising portion may be too short, compared with the decay, to be able to determine 
any trend in \epeak; this is common with weaker events, especially if there is only a single spectrum 
in the rising portion before the peak that is significant enough to obtain a fit. We take this as a motivation 
for our hypothesis: 
that signal-to-noise sampling under-emphasizes the lowest and highest energies, which are found preferentially 
on the rise and decay portions of pulses (true for both HTS and HIT pulses), resulting in the observed 
narrow distributions.

In this section, we will concentrate on bright, single pulses with HTS behavior. There is ample 
motivation for this: several very bright bursts are either composed of single pulses or have at 
least one well-separated pulse with clear HTS evolution. 
In several of these cases, the temporal evolution of \epeak\ is that of a $-1$ power law throughout. 
Since the picture becomes more complicated when physical processes are used to model the emission 
(in the case of \citet{burgess14,preece2014}, the sum of blackbody and synchrotron was used and the 
resulting temporal evolution of the characteristic energy was more consistent with a broken power 
law), we will restrict ourselves here to the evolution of the COMP function parameters.

\subsection{A Model Pulse with Spectral Evolution}

We start out with a simple simulation of a single pulse with HTS spectral evolution. Following 
\citet{2005ApJ...627..324N} and \citet{HakkilaPreece}, we adopt a parameterized pulse profile:
\begin{equation}
p(t) = \exp\left(-\tau_1/(t-t_{\rm start}) - (t-t_{\rm start})/\tau_2\right),
\end{equation}
where $\tau_1$ and $\tau_2$ are the rise and decay times of the pulse and the onset of the pulse is 
at time $t_{\rm start}$. In terms of these parameters, the duration of the pulse is approximately
\begin{equation}
t_{\rm dur} = \tau_2 \sqrt{9 + 12 \sqrt{\tau_1 / \tau_2}}.
\end{equation}
Next, we can specify a power-law temporal evolution for \epeak\ from $E_{\rm hi}$ to 
$E_{\rm lo}$ over the pulse duration $t_{\rm dur}$ by the power-law index:
\begin{equation}
i = \ln(E_{\rm hi}/E_{\rm lo})/\ln(t_{\rm dur}+ 1). \label{decay_index} %-t_{\rm dur}
\end{equation}
Finally, the power-law behavior of \epeak\ is expressed as:
\begin{equation}
E_{\rm peak}(t) = E_{\rm hi} \left(t -t_{\rm start} +  1\right)^{-i}.
\end{equation}
All of these quantities are so far continuous; we must discretize the time $t$ into finite width time bins. To 
simulate a pulse, we then evaluate a photon function for the changing value of \epeak\ in each of the time bins. 
In the following, we use the exponentially-attenuated cut-off power law form of Eq.\ \ref{comp}. This 
simplifies the analysis for the reasons discussed above: excluding the possible evolution of 
$\beta$, which could lead to complications in the analysis.
%, as the sum of Band functions is not necessarily a Band function. In 
%particular, this arises because the sum of changing power laws is not a power law. We investigate the 
%opposite behavior below, where the sum of cut-off power laws could be interpreted as a Band function. 
The photon function is converted to counts data by multiplication by a detector response matrix (DRM). 
For this to work, the energy bins of the photon function must match those of the specific DRM chosen. 
Different instruments will have different numbers of `input' photon energy bins, as well as `output' 
energy-loss bins on the counts side that match a specific instrument's data types. For this study, 
we use GBM DRMs from a triggered burst that have 128 output energy bins, corresponding to the CSPEC or TTE 
data types \citep{Meegan}. The result is a realization of a single pulse that consists of a 2D set 
of numbers that represents count rates in time and energy.

To simulate the correct counts statistics of an observed burst, we first scale the rates to produce 
some desired peak intensity. This peak count rate is multiplied into the entire 2D pulse model. 
We then add the background rate (as described below) to obtain a total model rate. Finally, we 
simulate the observed counts statistics by multiplying the total rates by the bin width in time to 
get total model counts for each time and energy bin. The counts are resampled randomly according to 
the Poisson distribution to obtain simulated observed counts. The total number of counts in each 
time bin determines the deadtime according to the GBM electronics model \citep{Meegan}. Together with 
the pulse model and background counts, these are saved in a FITS file that mimics actual GBM data. 
Note that \citet{BasakRao12} have a similar procedure of pulse spectral evolution simulation, 
used for fitting burst lightcurves and spectra simultaneously. They used the \epeak\ evolution model 
of \citet{LiangKargatis96}.

In real data, there is a certain level of background 
noise rate in each energy channel, usually changing (hopefully slowly) as a function of time. 
Modeling the background is important as it sets the scale of the signal versus noise for the 
entire simulation. Since our pulse model has an exponential rise and decay in time, it will be 
important to determine which spectra from our simulation have enough total counts to be usable 
for spectroscopy and which spectra may have to be added together (binned) with others in order 
to meet a preferred level of significance. In order to reduce the variance (especially important for 
low count spectral bins at higher energies), we start with a long background accumulation from 
actual burst data. Following our procedure for the BATSE and GBM spectroscopy catalogs \citep{KanekoCat,GoldsteinCat,GruberCat} 
we fit a low-order ($\le$ 4) polynomial to the temporal variations of the background 
and interpolate the fitted model during the burst interval. Usually, we select regions for background 
fitting both before and after the burst, to avoid the need to extrapolate the background model.
Since the background is fit to a polynomial in time in each energy bin, the variance associated with 
each bin is governed by the Gaussian statistics of the background temporal model parameters 
(the fitted polynomial coefficients). This variance is recovered in our simulations by resampling 
the background counts from a normal distribution, given the background model error for that bin. 

\begin{figure*}
\epsscale{2.35}
\plottwo{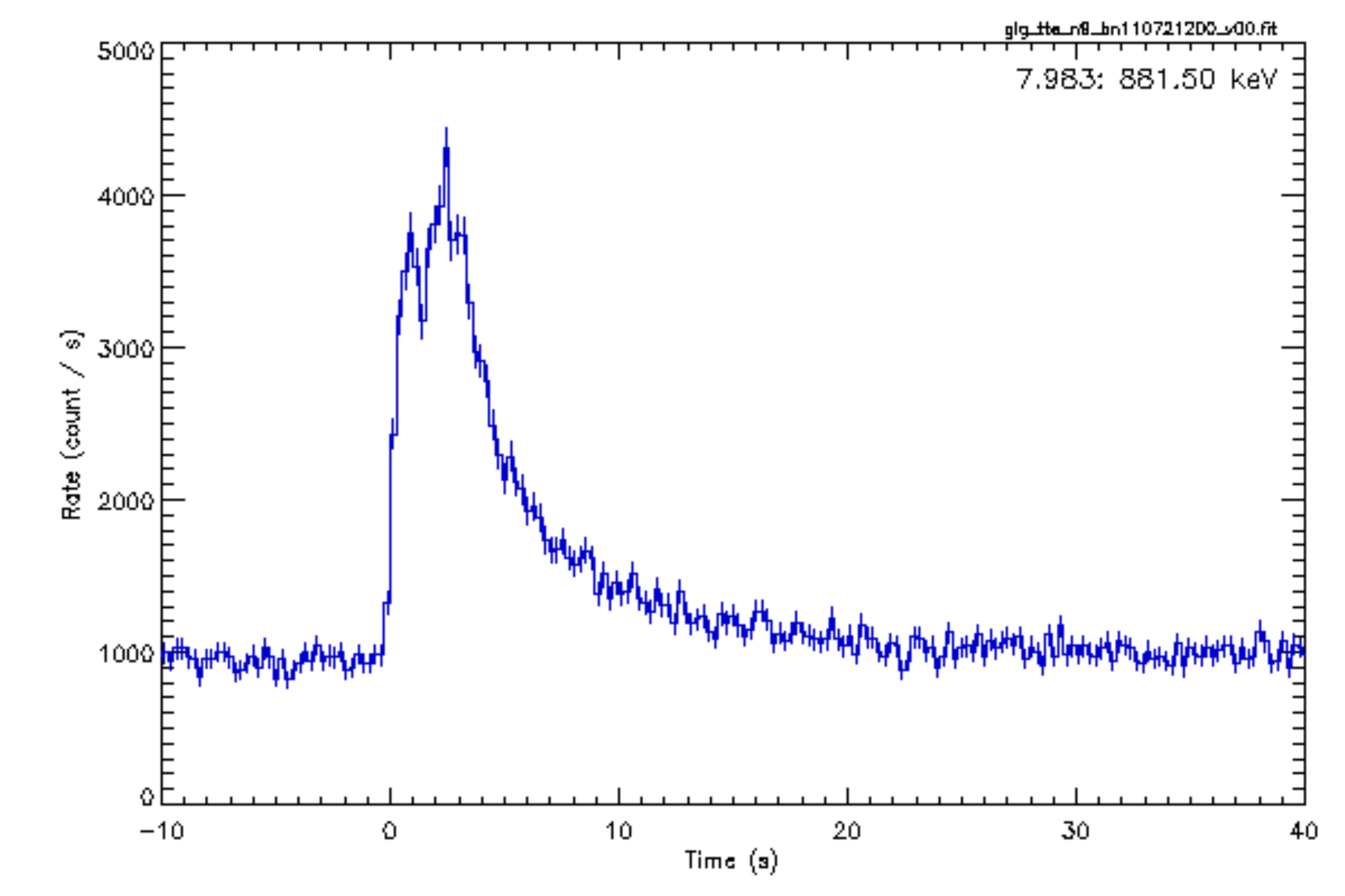}{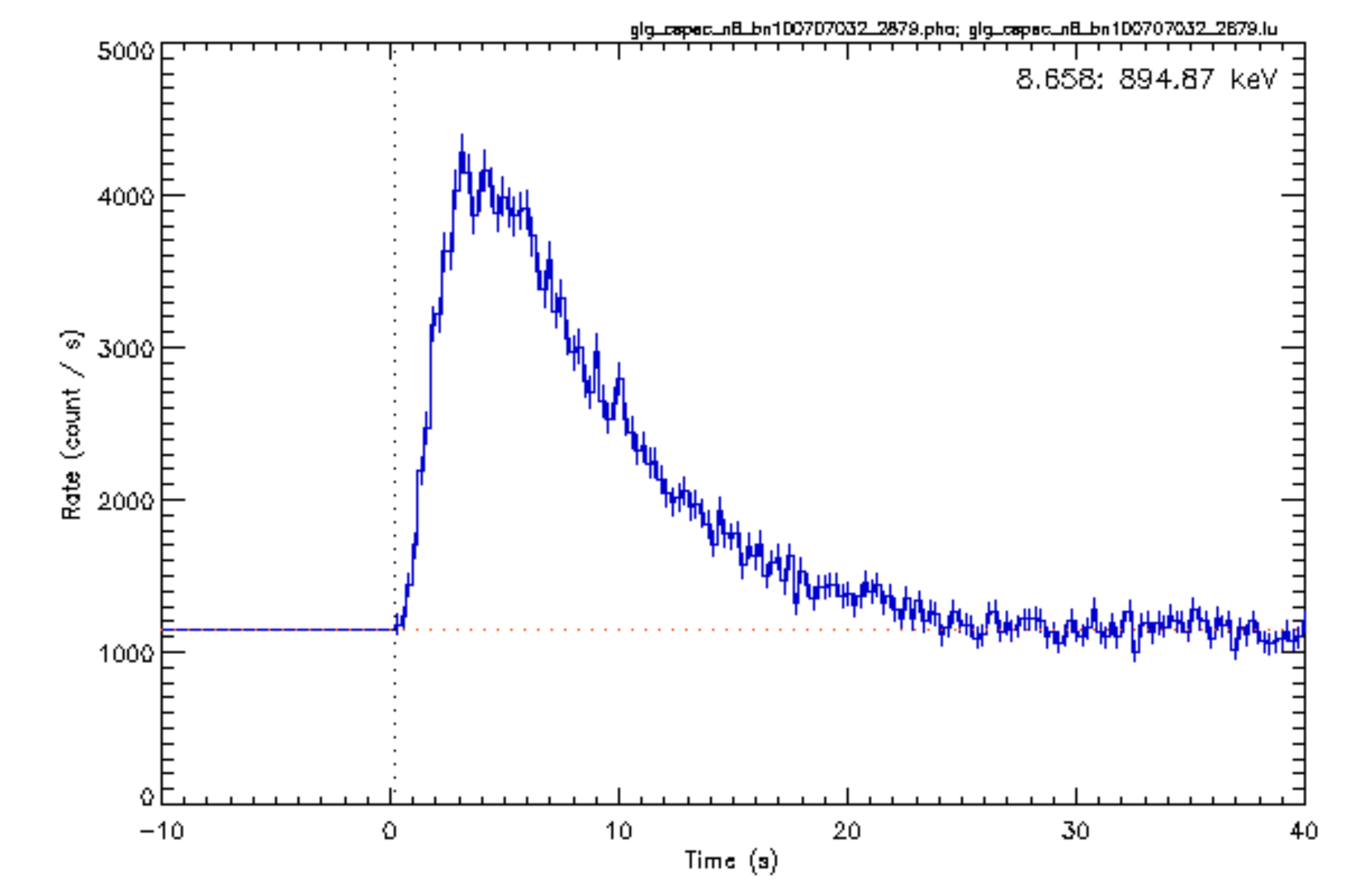}
\caption{The GRB 110721A lightcurve from GBM NaI detector 9 at 256 ms time resolution shows some deviation 
from a simple pulse profile at the peak (left). 
Compare with a simulation derived from the Norris pulse model, with 3 s rise and 5 s decay time constants  
and 3000 count/s peak count rate scaling above background (right). The two figures are presented with the same x-axis limits; 
however no attempt was made to optimise the pulse parameters to make the simulated pulse profile match 
with the GBM data. \label{htslc}}
\end{figure*}

\begin{figure}[h]
\includegraphics[scale=0.5]{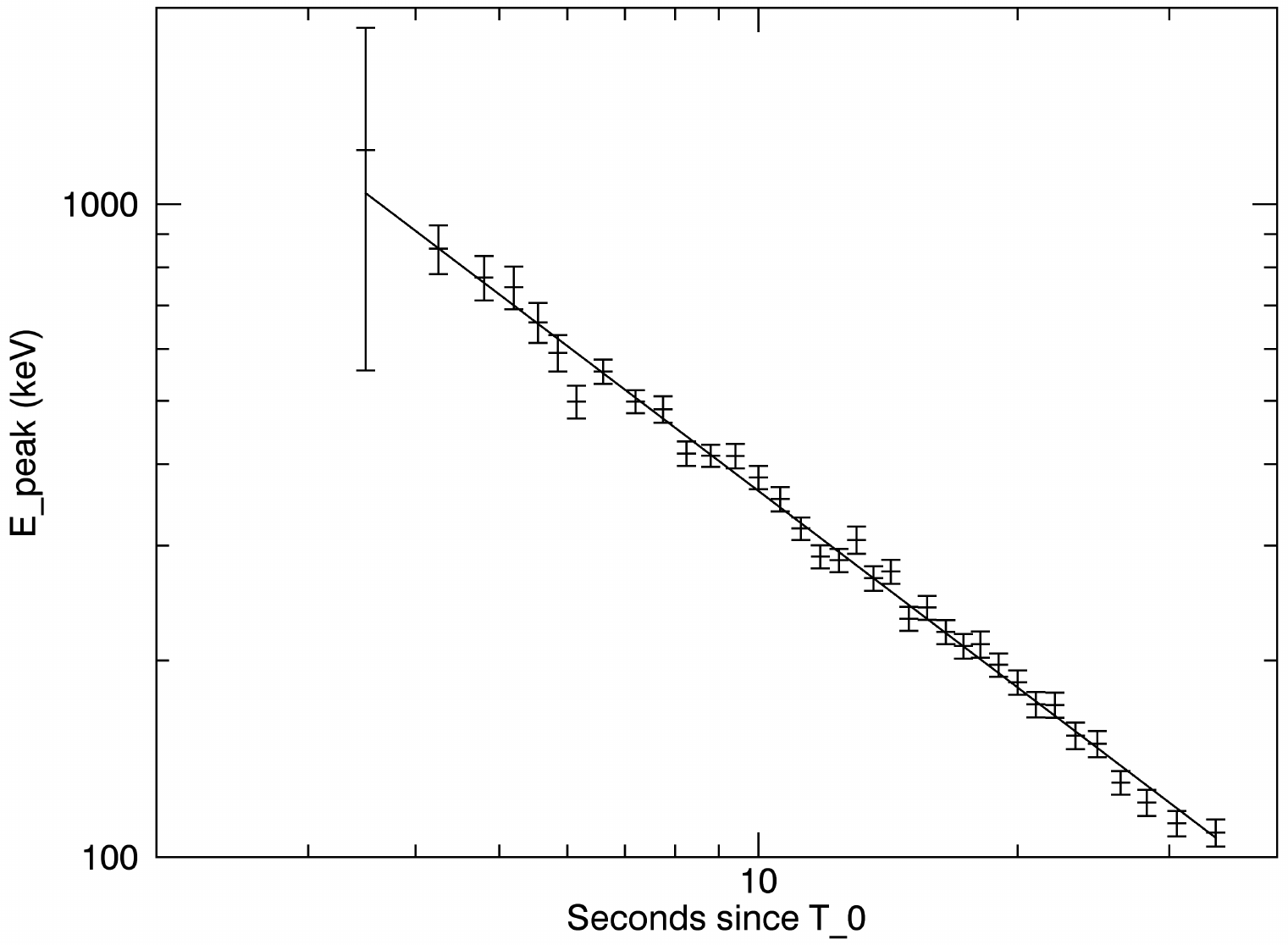}
\caption{Fitted \epeak\ values from an example pulse simulation shows that the input $-1$ power law 
spectral evolution is recovered after signal-to-noise temporal binning and forward-folding with the 
detector response. In this case, the pulse parameters were 3 s rise time, 11 s decay time, scaled to 
40,000 count/s at the peak. \epeak\ evolves from 1500 keV to 100 keV over the nominal duration of 
the pulse. \label{Sim_EP_Fit}}
\end{figure}

The resulting simulated pulse is remarkably simple, yet it reproduces a number of characteristics 
of observed single pulses. First of all, the light curve clearly mimics the observations, as 
expected, given that the Norris pulse model was designed for this purpose (e.g.\ Figure \ref{htslc}). We should mention that 
a pulse stacking analysis by \citet{hakkilapreece14} has shown that significant residuals exist on top of the Norris pulse 
shape that do not represent separate pulses; for example, there is a shoulder after the peak in 
Figure \ref{htslc} at T+3 s as well as in the published GBM NaI light curve for GRB 130427A at roughly T+1.2 s. 
To keep the analysis simple, we do not include this effect here. Upon performing spectral fitting 
of each time bin, the simulated power-law 
\epeak\ evolution from hard to soft ($E_{\rm hi} = 1500$ keV and $E_{\rm lo} = 100$ keV) is 
recovered from the simulation, as can be seen in Figure 
\ref{Sim_EP_Fit}. In this simulation, the peak count rate was chosen to be extremely high, 
40,000 count/s, even so, during the rise and decay portion of the pulse some low-count time bins 
had to be combined to achieve a minimum signal to noise threshold of 45. This threshold ensures 
that there are enough total counts in the binned spectra to do useful spectroscopy, as described 
in the BATSE and GBM Spectroscopy Catalogs. 

We have checked the effect of the spectral parameter $\alpha$ by performing three identical simulations, 
varying only the $\alpha$ value in three steps between $-0.66$ and $-1.2$, bracketing the most likely fitted 
value of $-1$ from the distributions reported in the various spectroscopy catalogs. For the three values simulated, there was no change in
the mean value of \epeak\ for the resulting distributions to well within the uncertainty on the mean: 
$177 \pm 2$ keV (these and following simulations are based on $E_{\rm hi} = 3000$ keV and $E_{\rm lo} = 50$ keV). 
This is expected, as \epeak\ does not formally depend upon $\alpha$ (nor does it depend upon the Band  
parameter $\beta$, as long as $\beta < -2$). As discussed above, we do not present any simulations of Band 
functions, so the evolution of $\beta$ is not considered here.

\subsection{Spectral Lag and Pulse Width Evolution}

\begin{figure}[t]
\includegraphics[scale=.45]{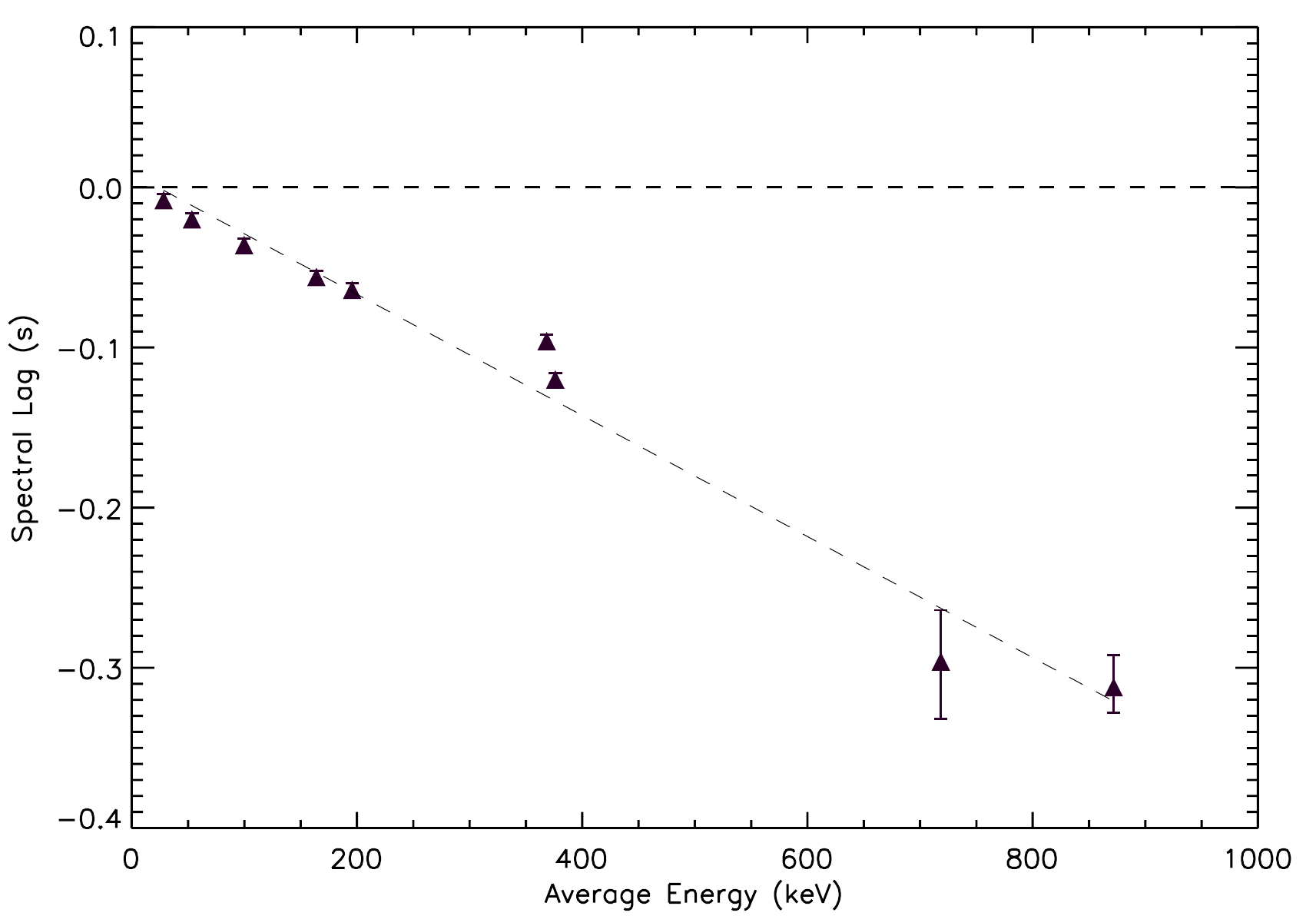}
\caption{The spectral lag in seconds for the simulated pulse in Figure \ref{Sim_EP_Fit} is plotted against the average 
energy of 9 spectral data bins of roughly equal width in log energy. Some of the BGO detector bins overlap in energy with
those of the NaI, which accounts for the odd spacing. The binning is the same as can be seen in Figure 
\ref{Rob_FWHM_Energy}. The lowest energy 
bin serves as the reference against which the lag of the higher energy bins is calculated and thus 
has zero lag by definition. \label{Rob_energy_dep_synthetic_Suzanne}}
\end{figure}

\begin{figure}[t]
\includegraphics[scale=.45]{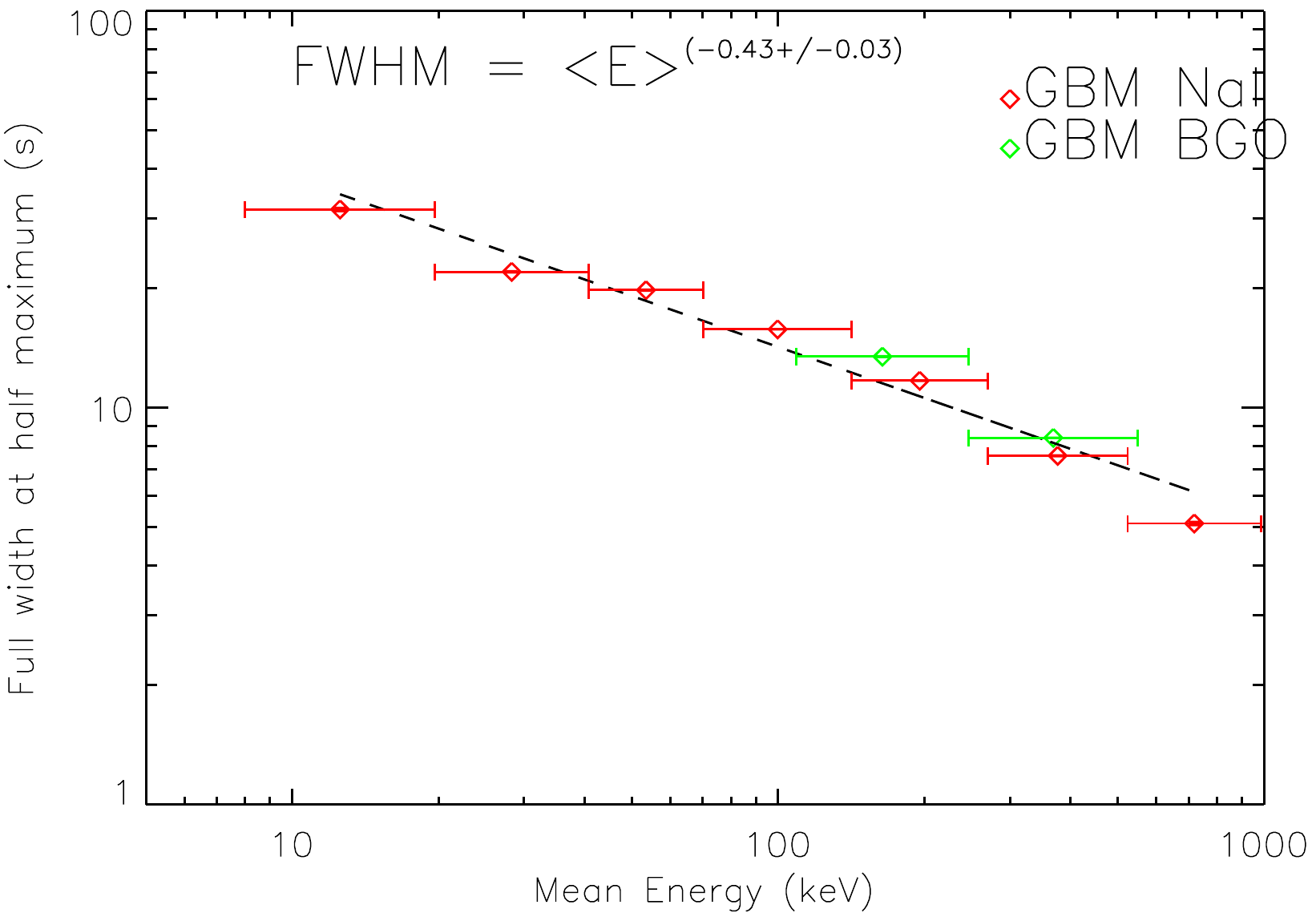}
\caption{The pulse widths in different energy bands in seconds for the simulated pulse in Figure \ref{Sim_EP_Fit} is plotted against the average 
energy of each of the 9 spectral data bins of roughly equal width in log energy. The trend of wider pulse width 
for lower energy is fitted with a power law, as suggested by \citet{2005ApJ...627..324N}. \label{Rob_FWHM_Energy}}
\end{figure}

\begin{figure}[ht!]
\includegraphics[scale=.25]{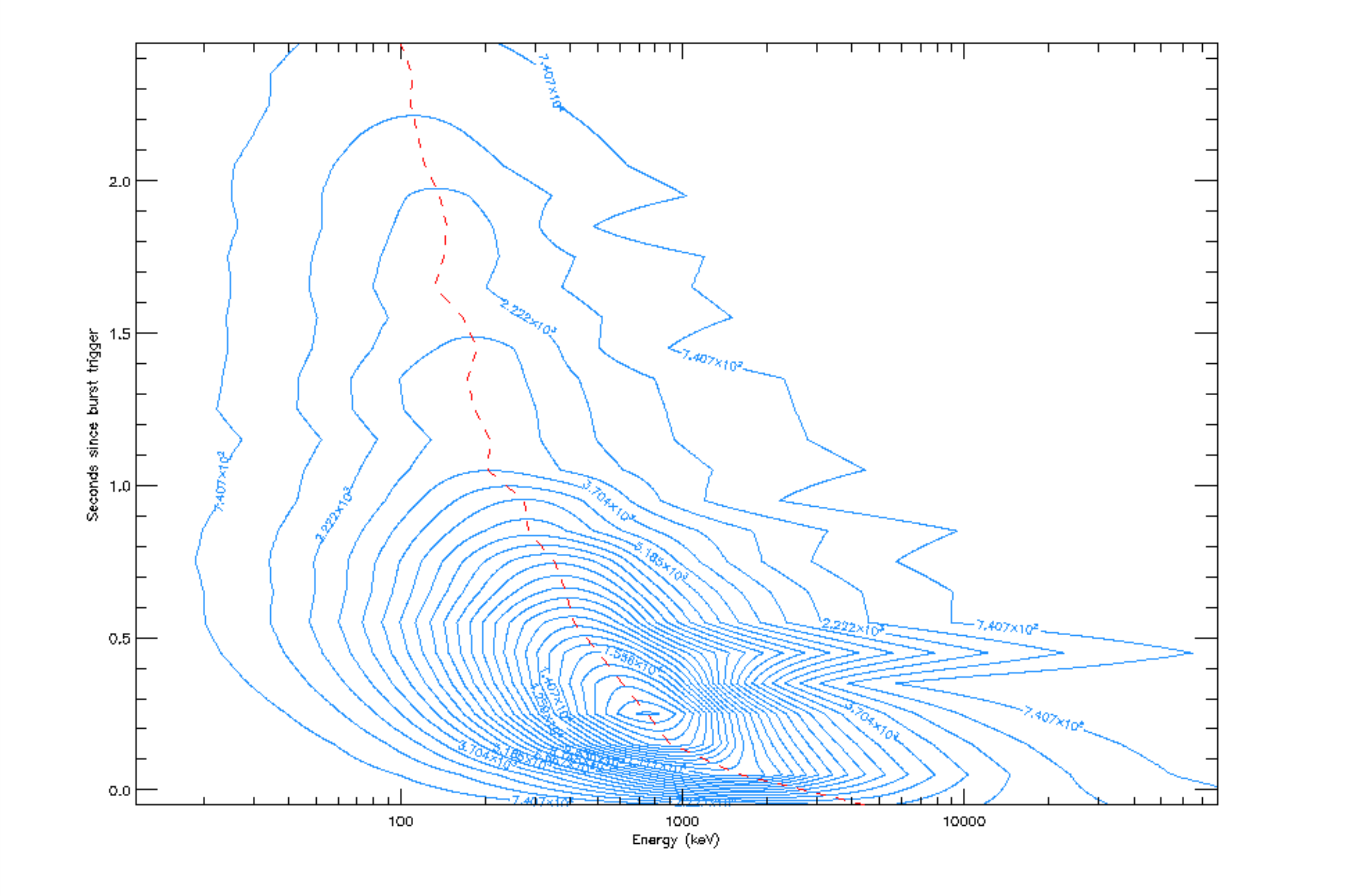}
\caption{The spectral evolution of fitted \epeak\ values during the first 3 s of GRB130427A is presented 
as the `continental divide' (dashed line) on top of a $\nu {\cal F}_{\nu}$ contour plot. Because \epeak\ 
represents the peak of each $\nu {\cal F}_{\nu}$ spectrum (arranged in time from bottom to top), 
the levels of increasing intensity form closed contours. The peak intensity clearly comes later 
than the highest \epeak, which is found at the start of the burst. \label{GRB130427A_Horiz_Contour}}
\end{figure}

This simulation of the temporal evolution of spectral parameters allows us to compare it with certain 
other properties of observed pulses, such as the spectral lag and pulse-width evolution as a function 
of energy, as characterized in the initial pulse of GRB 130427A \citep{preece2014}.  
Because the simulations and the spectral fits share the same power law evolution model, the simulation reproduces the 
lag-width relation of the real data, when the lag is taken with respect to a fixed low energy channel. 
That is: the spectral lag increases as a function of increasing 
energy (\citet{KocevskiLiang}; see Figure \ref{Rob_energy_dep_synthetic_Suzanne}). In addition, the pulse width decreases 
with increasing energy (\citet{richaga}; see Figure \ref{Rob_FWHM_Energy}). Except for the factor of ten increase in the 
timescales associated with the simulations, these results are qualitatively consistent with the 
results presented in Figure 1 in \citet{preece2014}. One can visualize how this behavior arises by 
examining the joint temporal-spectral model as a contour plot. In Figure \ref{GRB130427A_Horiz_Contour}, 
we compare the fitted data from GRB130427A with the fitted data from our simulated pulse. Log energy 
is on the horizontal axis, while time is along the vertical. The energy of the peak in $\nu F_{\nu}$ 
(i.e.: \epeak) is traced over time by the dashed red line and can clearly be seen to exhibit a hard-to-soft 
trend, as indicated by movement toward the left in the plot. A vertical cut or band corresponds to a 
specific energy. The pulse peaks appear later in the lower energy bands, which is the source of the 
spectral lag. At the same time, the temporal power-law nature of \epeak\ ensures that an increasing number of 
spectra with similar \epeak\ values are found in a given accumulation time with respect to similar 
accumulations at the start of the pulse, where the spectra are varying more rapidly. This is the source 
of the increase in pulse temporal width with lower energy. It is interesting to speculate how this 
might be extended in the opposite direction in time with respect to the highest peak energies. 
The observed trend is that the pulse width becomes increasingly narrow with higher energies at the same time that 
the pulse peak lag gets shorter, so that the temporal width approaches zero. 
The limiting result is that the highest energy photons are all confined to a very short time window very 
close to the GBM trigger time. Of course, real detectors 
are limited to counting photons, where the continuum model must break down. We note that there is a cluster 
of 3 Fermi LAT high-energy ($> 100$ MeV) photons in a 0.1 s window at or just before the GBM trigger time of GRB130427A, 
followed by a gap of 5 s.

\begin{figure}[ht!]
\includegraphics[scale=.45]{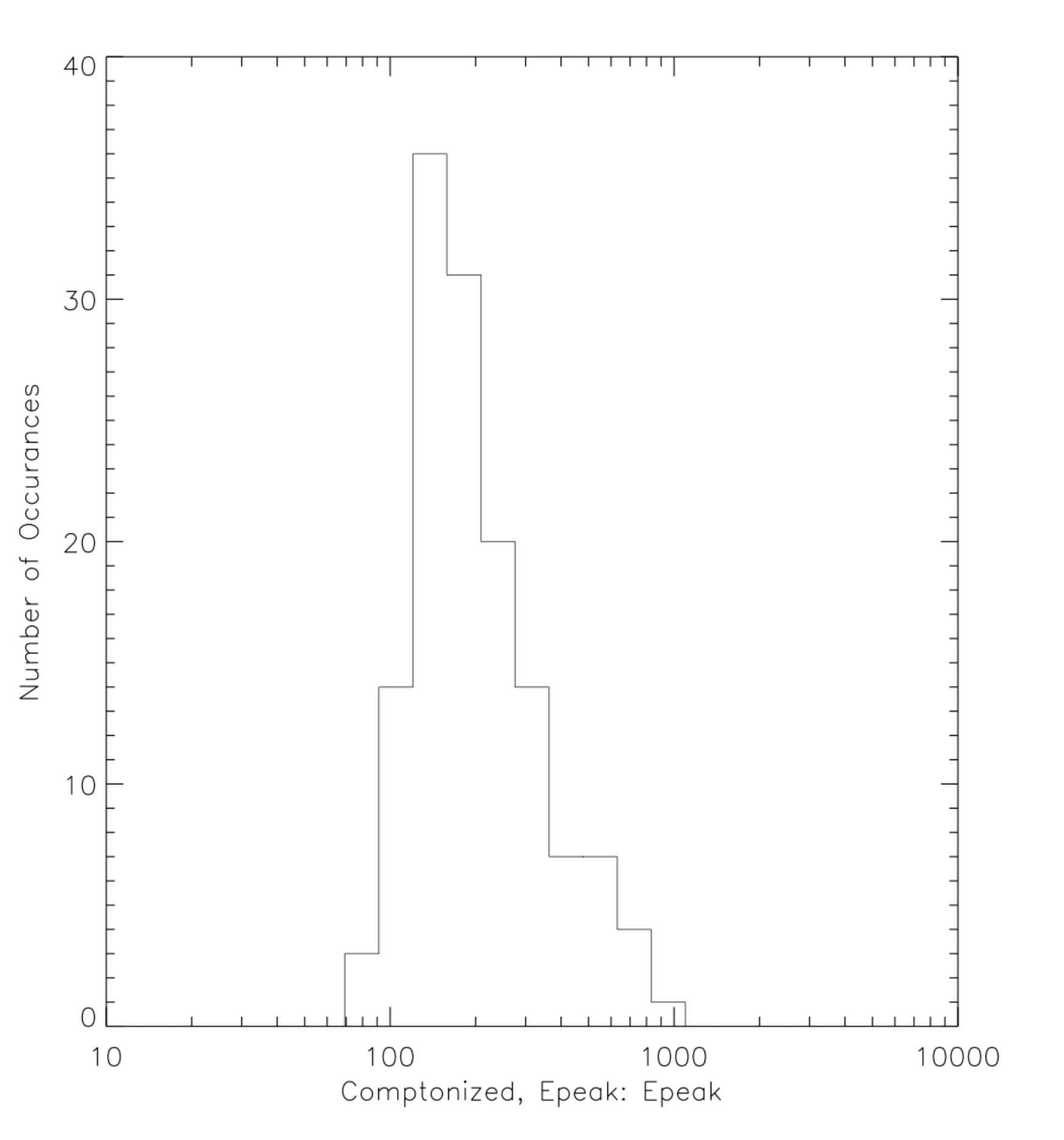} %{spec_mdl_BandsGRBEpeakHisto_v01.png}
\caption{The distribution of fitted \epeak\ values for a simulation of a single, bright GRB pulse 
that has strong spectral evolution is shown as a histogram. Although the pulse is well sampled throughout 
its time history, the predominance of samples occurs at the peak of the pulse, heavily weighting the 
average spectrum with medium-energy \epeak\ spectra. \label{ephisto}}
\end{figure}

\begin{figure}[h!]
\includegraphics[scale=.45]{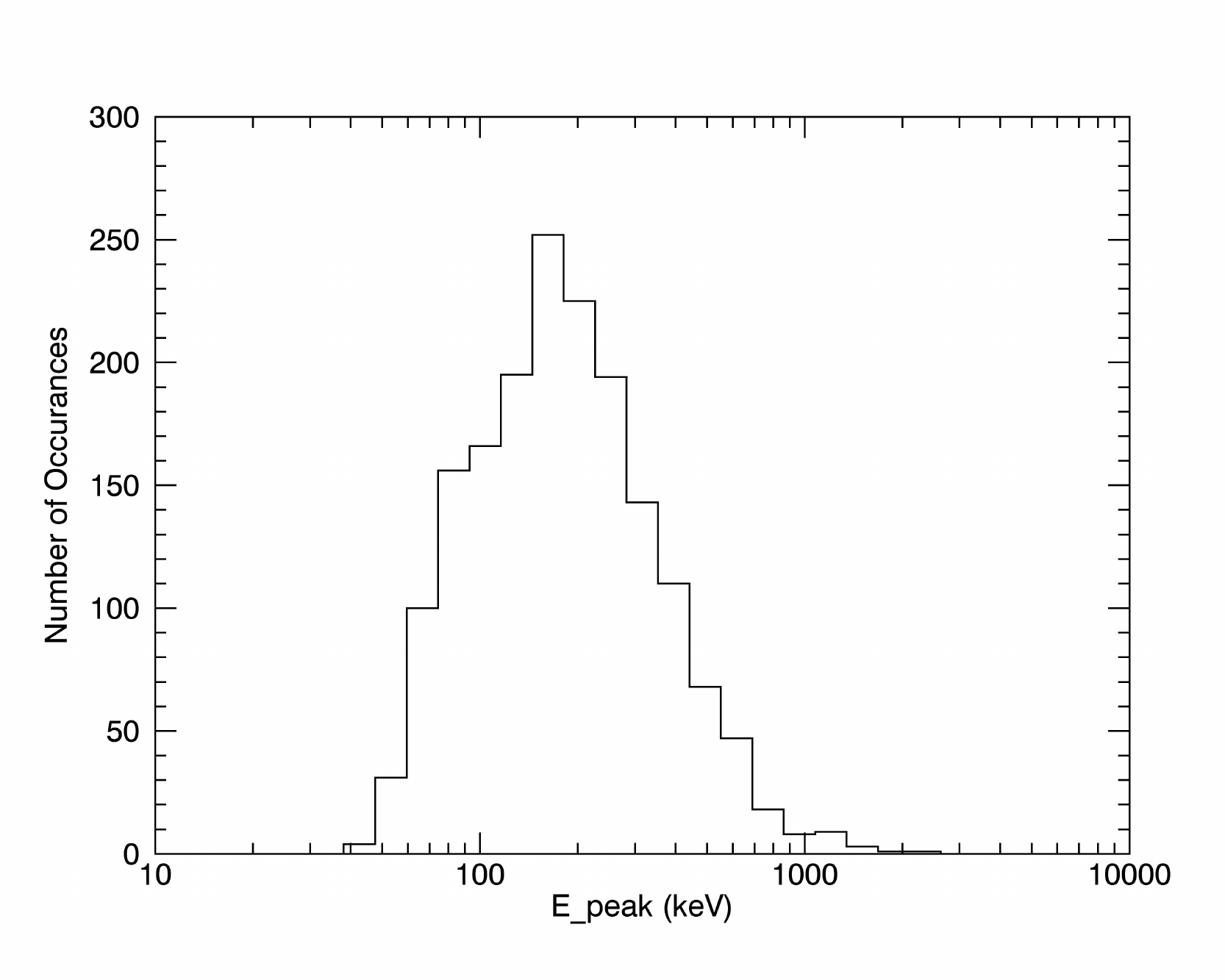}
\caption{The distribution of fitted \epeak\ values is plotted as a histogram for fifty simulated single-pulse GRBs of ten 
different brightnesses and five different pulse asymmetries. Each simulated lightcurve has been binned 
to the same signal to noise ratio, resulting in different binning for each, while retaining similar 
statistics for the spectral fitting. The histogram represents the ensemble of all time-resolved spectra, 
similar to the BATSE or GBM Spectral Catalogs of Bright bursts, to be compared with Figure \ref{BATSE_GBM_Epeak}.
\label{sim_ep_histo}}
\end{figure}

The high intensity of this initial simulation allows us to examine the results with high temporal resolution. 
In particular, we can bin the fitted \epeak\ values according to number of occurrences in log energy, 
the result is shown in Figure \ref{ephisto}. We have simulated a larger sample of 
spectra by using the time-resolved spectral fits from fifty simulations of bursts of varying 
brightness and different rise and decay parameters (Figure \ref{sim_ep_histo}). The spectra were drawn from the 
set of simulations described below in Section \ref{asymsection}, all binned to the same signal-to-noise criterion. 
Thus, the brightest simulated bursts contributed the most spectra at the highest temporal resolutions, 
while the dimmest bursts contributed few spectra, taken with relatively coarser binning. Clearly, as 
with the dimmer bursts from the actual observations, the longer temporal accumulations can average 
over significant spectral evolution in a single spectrum.

\begin{figure}[t!]
\includegraphics[scale=.45]{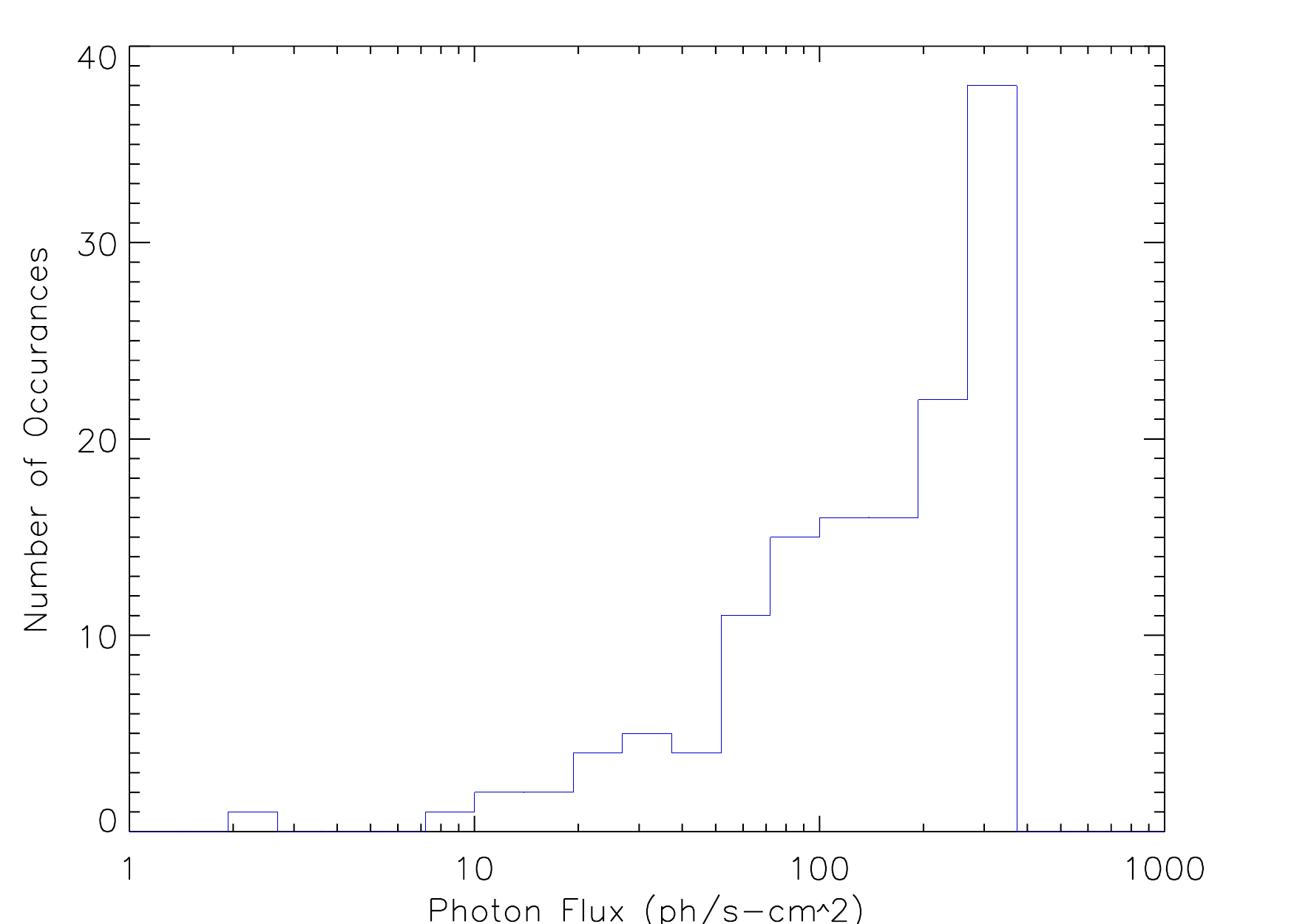}
\caption{The distribution of photon flux values is displayed as a histogram for the bright single-pulse 
simulation shown in Figure\ \ref{ephisto}. The single most populated bin corresponds with the peak in the 
lightcurve, which, by having the highest SNR, is the most frequently sampled portion of the lightcurve. 
\label{sim40k_phot_flux_histo}}
\end{figure}

The pulse histogram in Figure \ref{ephisto} clearly has the 
largest contribution from the portion in the lightcurve that is the most densely sampled; that is: from the  
region around the peak in the light curve, where \epeak\ doesn't change much. This is a sampling bias 
that can be seen even more dramatically in a histogram of the photon flux (Figure\ \ref{sim40k_phot_flux_histo}), which is sharply 
concentrated at the flux values near the peak of the lightcurves. Where \epeak\ is highest and lowest, the 
lightcurve is rapidly changing during the rise and decay portions, respectively, and thus the sampling of \epeak\ values is 
relatively sparser than at the peak of the lightcurve.  

\subsection{The Effect of Evolution in $E_{\rm hi}$}

So far, the simulations have only modeled one specific case of spectral evolution: they were all based 
upon the fixed values $E_{\rm hi} = 1500$ keV and $E_{\rm lo}=100$ keV, resulting in the same value for the temporal 
decay index $i\approx -1$ (Eq.\ \ref{decay_index}). Either one or both of these may naturally be considered 
to be drawn from a fairly large set of values, due to the spread in redshift and bulk Lorentz factors, 
as discussed in the Introduction (Eq.\ \ref{e_obs}). We next investigate how these parameters influence 
the \epeak\ distributions. 

First, we generated pulse simulations by varying both $E_{\rm hi}$ and $E_{\rm lo}$ by the same factors of two: %E_HI -> 2X E_HI and E_LO
$(1/4,1/2,1,2,4)$. This has the effect of keeping the value of the temporal decay index $i$ the same. The result could be predicted:
the mean value for the resulting \epeak\ histograms for each pulse also change by the same factors of two. This 
behavior holds over the entire range of the inputs $E_{\rm hi} = (750; 1,500; 3,000; 6,000; 12,000)$ keV (5 data points). 
This range was determined mainly by its effect on $E_{\rm lo} = (12.5; 25; 50; 100; 200)$ keV:
anything below 12.5 keV is not recoverable and anything above 200 keV is already too high to be very 
common, based upon the observed behavior that bursts with average or peak \epeak\ values greater than 1 MeV 
are rare. In any case, the linearity in the peak \epeak\ from this simulation suggests that one could 
build up any \epeak\ distribution whatsoever by weighting pulses with appropriate energy limit parameters. 
The most natural distribution prior for both $E_{\rm hi}$ and $E_{\rm lo}$ may be uniform, in which case, we 
should expect that the \epeak\ distributions be more flat-topped than they appear to be. 

\begin{figure}[t]
\includegraphics[scale=.35]{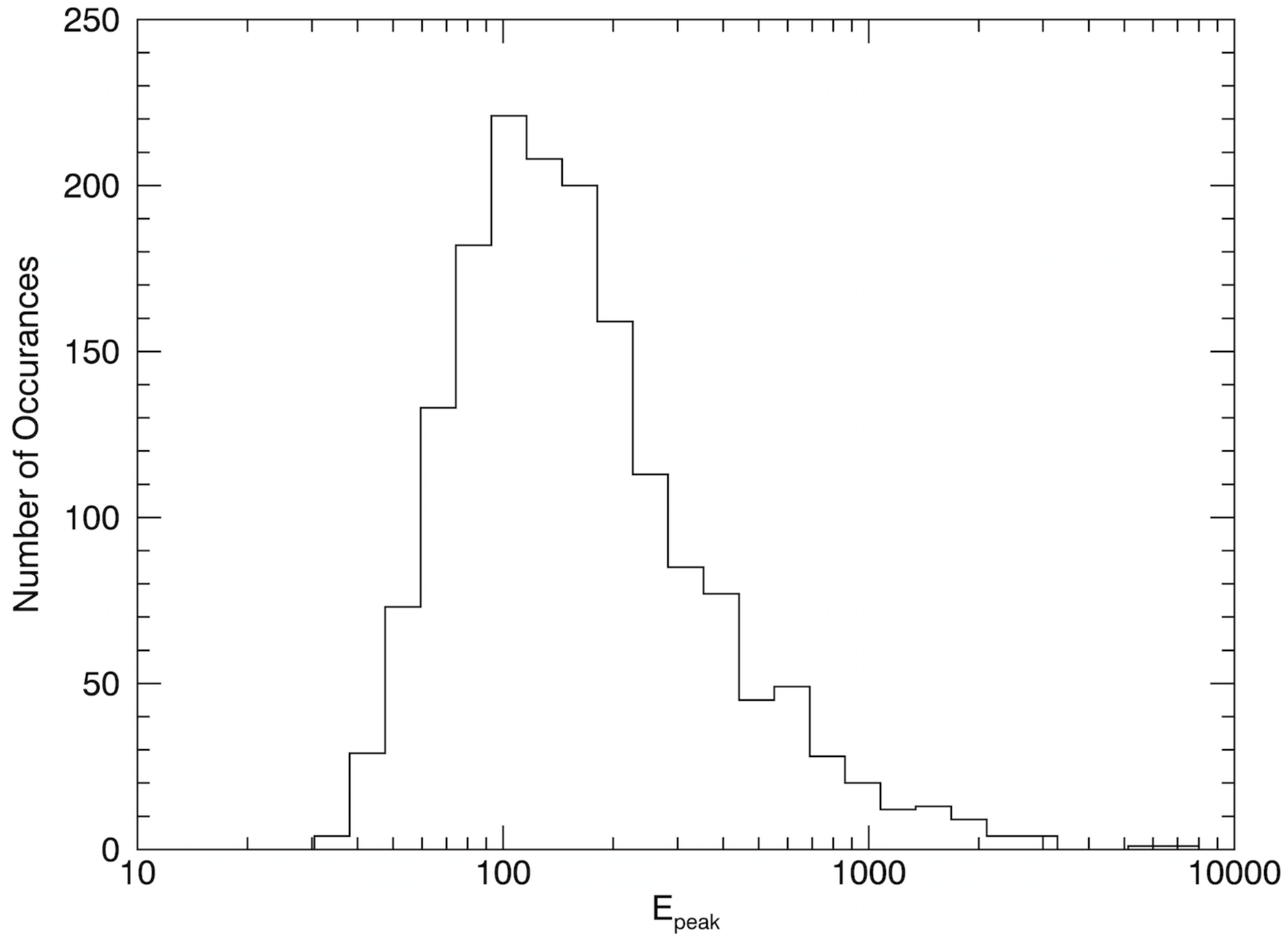}
\caption{The distribution of fitted \epeak\ values is plotted as a histogram for six simulated 
single-pulse GRBs only differing in their values of $E_{\rm hi}$. 
\label{sim40k_E_HIEvo_dist}}
\end{figure}

Next, we varied only $E_{\rm hi}$ by factors of two and kept $E_{\rm lo}$ constant. The corresponding 
change in the temporal decay index can be seen in Table \ref{ehi_evo}. Here, we created 6 individual 
bright pulses with parameter values from $E_{\rm hi} = 12,000$ to $375$ keV.
The mean \epeak\ values of the distributions do not change much: 270 to 93 keV,
which is roughly a factor of 3 for a change by a factor of 32 in $E_{\rm hi}$. Apparently, a large
change in $E_{\rm hi}$ leads to only a modest shift in the \epeak\ distribution; the behavior of $\log(E_{\rm hi})$ 
versus $\log(\left<E_{\rm peak}\right>)$ is definitely linear. As an aside, we could change the pulse parameters 
in each case to achieve a constant $i=-1$ decay index, but as these bright pulses are quite well sampled, it 
would make little to no difference in the results. A histogram of all the fitted \epeak\ values from these 6 
simulated pulses is slightly skewed toward the low energy side, as seen in Figure \ref{sim40k_E_HIEvo_dist}. This arises because 
the distributions with lower $E_{\rm hi}$ are narrower, while they have roughly the same number of spectra. 
Despite this unbalanced weighting, the overall distribution still peaks at a median value, and is no 
more flat topped than Catalog-derived distributions.

\begin{table}[t!]
\begin{center}
\begin{tabular}{r c c}
\tableline\tableline
$E_{\rm hi}$ &  $\left<E_{\rm peak}\right>$ &  $i$ \\
         keV &          keV                & Decay Index \\
\tableline
$12,000$ & $272 \pm 3$ & -1.45 \\
$6,000$ & $218 \pm 2$ & -1.16 \\
$3,000$ & $177 \pm 2$ & -1.08 \\
$1,500$ & $143 \pm 2$ & -0.9 \\
$750$   & $115 \pm 2$ & -0.71 \\
$375$   & $93 \pm 1$ & -0.53 \\
\tableline
\end{tabular}
\caption{Results of simulations of bright, long pulses with different $E_{\rm hi}$, but holding $E_{\rm lo}=50$ keV constant. 
The mean value of \epeak\ for each simulated time sequence is given by $\left<E_{\rm peak}\right>$. 
The temporal decay index is calculated from Eq. \ref{decay_index}.\label{ehi_evo}}
\end{center}
\end{table}

Which scenario may be preferred here depends upon what one expects as being more reasonable: is the
temporal decay index of $i=-1$ a physical value, so we should scale both $E_{\rm hi}$ and $E_{\rm lo}$ by the same
factor? We don’t have a justification for this, but we have two examples of observed bursts with
roughly this same temporal index (GRB110721A and GRB130427A): the durations scale in just such a manner
to make this work. On the other hand, in the previous section, we put forth the
hypothesis that pulses are initiated by an impulsive energization event that loses energy over
time. In which case, only the initial $E_{\rm hi}$ value is important. Certainly, the maximum fitted
\epeak\ values for the two events are different by a factor of three or so. The value of $E_{\rm lo}$
would then be determined by the last point at which spectral fitting can be done on the
decaying portion of the burst, not by any physical constraint. At the same time, a physical argument 
may support the concept of a conserved temporal decay index for this scenario as well: in order to 
maintain this index, the duration would certainly have to be coupled to the energy loss mechanism. 

\subsection{Spectral Characteristics of the Pulse-Averaged Spectrum}

\begin{figure*}[ht!]
\plottwo{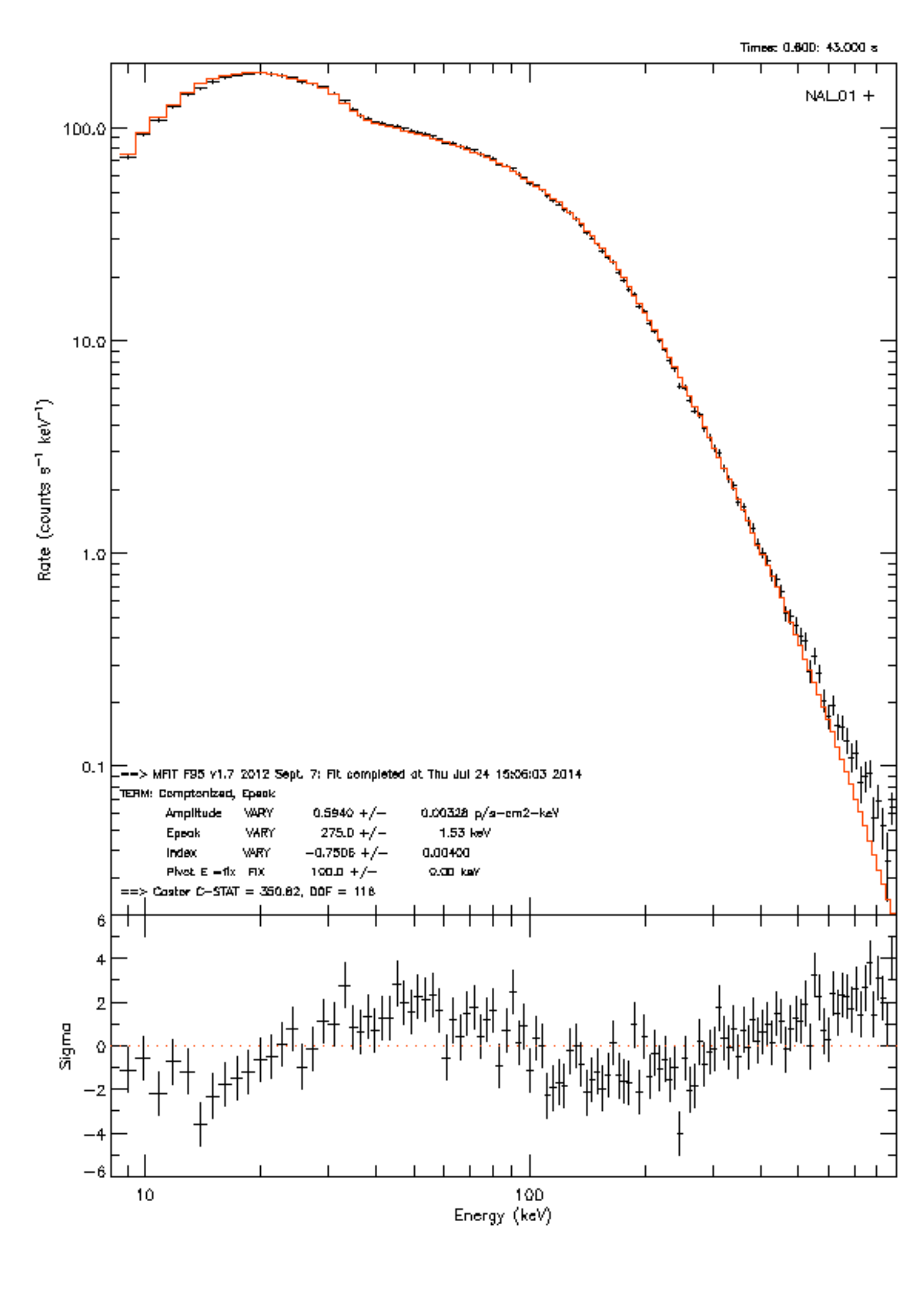}{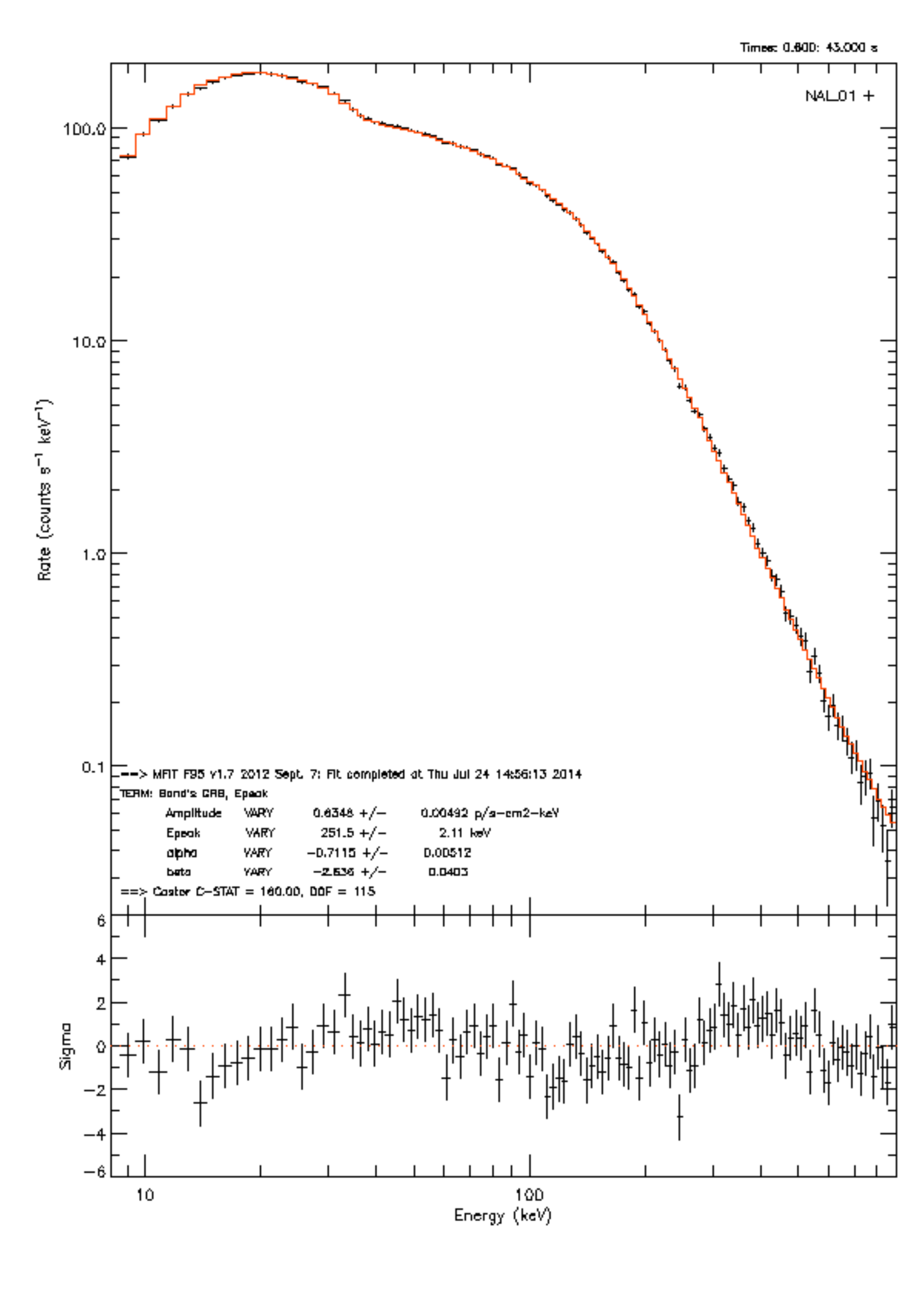}
\caption{Time averaged spectral fits to the simulation of a very bright single-pulse GRB with strong 
spectral evolution (as described in Figure \ref{Sim_EP_Fit}). The average spectrum is fitted with the 
input spectral function (cut-off model power law -- COMP, {\em left}) as well as the Band function 
({\em right}). The residuals plots at the bottom of the left-hand figure reveals 
strong discrepancies with the assumed COMP model that are largely absent in the Band function fit. The 
change in C-STAT between the two fits is 191 for one additional degree of freedom, suggesting that the 
sum of several COMP functions with strong evolution in \epeak\ and intensity do not preserve the COMP 
shape.
\label{COMPFitSpectralDistortions}}
\end{figure*}

There are two things to note concerning the spectral fit to the average spectrum obtained by summing 
all the individual spectra in the simulated pulse. The first is that 
although the simulation was constructed with the cut-off power law function, the spectral evolution 
imposed throughout the pulse conspires to deform the average spectrum significantly. In fact, the 
pure cut-off power law function does not result in an acceptable fit, as evidenced by the significant 
runs of residuals to the spectral model, as seen in Figure \ref{COMPFitSpectralDistortions} ({\em left}). As it 
happens, a fit to the same average spectrum using the Band function is much improved (CSTAT = 160, 
compared with CSTAT = 351, which is a considerable improvement for one degree of freedom; see Figure 
\ref{COMPFitSpectralDistortions}, {\em right}). Thus, at least for the HTS spectral evolution we have 
simulated, the sum of exponentially cut-off spectra can result in an average spectrum that is 
consistent with the Band function, which incorporates a high-energy power-law. 

The other point is that the sampling bias shown in Figure \ref{sim40k_phot_flux_histo} implies that values for 
\epeak\ derived from a fit to the average spectrum of a single-pulse burst should be very well correlated 
with that of the peak spectrum. The results for 11 single-pulse GRBs are shown in Table \ref{avevpeak}. 
The Pearson's rank correlation between the average and peak \epeak\ values is 0.77, which has a chance 
probability of 0.005. The `peak' \epeak\ is given by $E_{\rm p,p}$ and the average spectrum 
(or `fluence') \epeak\ is $E_{\rm p,f}$; both are in keV.
In the last column in the Table, the difference between the peak and average spectra \epeak\ values is 
expressed in units of the r.m.s.\ errors of the two values. Not only are the peak values always greater 
than the average, but nearly every pair differ by less than three standard deviations. For asymmetric 
pulses that peak early, the number of spectra after the peak in the lightcurve outnumber those before 
it, so a fluence-weighted spectrum will be dominated by lower \epeak\ spectra, compared with that of 
the peak spectrum itself, as discussed below. This accounts for the peak \epeak\ being greater than the fluence \epeak.

\begin{table}[t!]
\begin{center}
\begin{tabular}{l c c c}
\tableline\tableline
GRB & $E_{\rm p,p}$ & $E_{\rm p,f}$ & Sigma Dev. \\
\tableline
081110601 & $192 \pm 49$ & $46 \pm 12$ & 2.8 \\
081224887 & $934 \pm 217$ & $361 \pm 30$ & 2.6 \\
090719063 & $361 \pm 39$ & $207 \pm 8$ & 3.8 \\
090809978 & $227 \pm 40$ & $158 \pm 10$ & 1.6 \\
100612726 & $143 \pm 10$ & $111 \pm 4$ & 2.7 \\
100707032 & $597 \pm 63$ & $239 \pm 10$ & 5.6 \\
110407998 & $779 \pm 309$ & $543 \pm 106$ & 0.6 \\
110721200 & $1506 \pm 756$ & $237 \pm 32$ & 1.7 \\
110817191 & $345 \pm 40$ & $207 \pm 16$ & 3.2 \\
110920546 & $902 \pm 279$ & $215 \pm 5$ & 2.5 \\
130427324 & $812 \pm 19$ & $784 \pm 9$ & 1.3 \\
\tableline
\end{tabular}
\caption{Comparison between Peak and Fluence \epeak. Column 1 lists the burst name in the GBM convention 
(yymmddfff: see \citet{GBMBurstCat}). Columns 2 and 3 show the fitted \epeak\ value in keV for the peak flux 
and the fluence spectra, respectively. The last column shows the difference between the two \epeak\ values 
in units of the root mean square of the two error values.\label{avevpeak}}
\end{center}
\end{table}

%Another way to see this is to stack all the deconvolved 
%$\nu {\cal F}_{\nu}$ spectra from a strongly evolving pulse on top of each other and overplot their 
%weighted average, as shown in Figure \ref{StackCOMP_b1} (dashed line). The weighted spectrum is 
%constructed by adding up each of the time-resolved spectra, weighted by the total counts in each. 
%This artificially constructed spectrum strongly resembles the deconvolved time-averaged spectrum and 
%has several features that are qualitatively different than any one of the time-resolved spectra that 
%it is made up from. First of all, 

\begin{figure}[t]
\includegraphics[scale=.45]{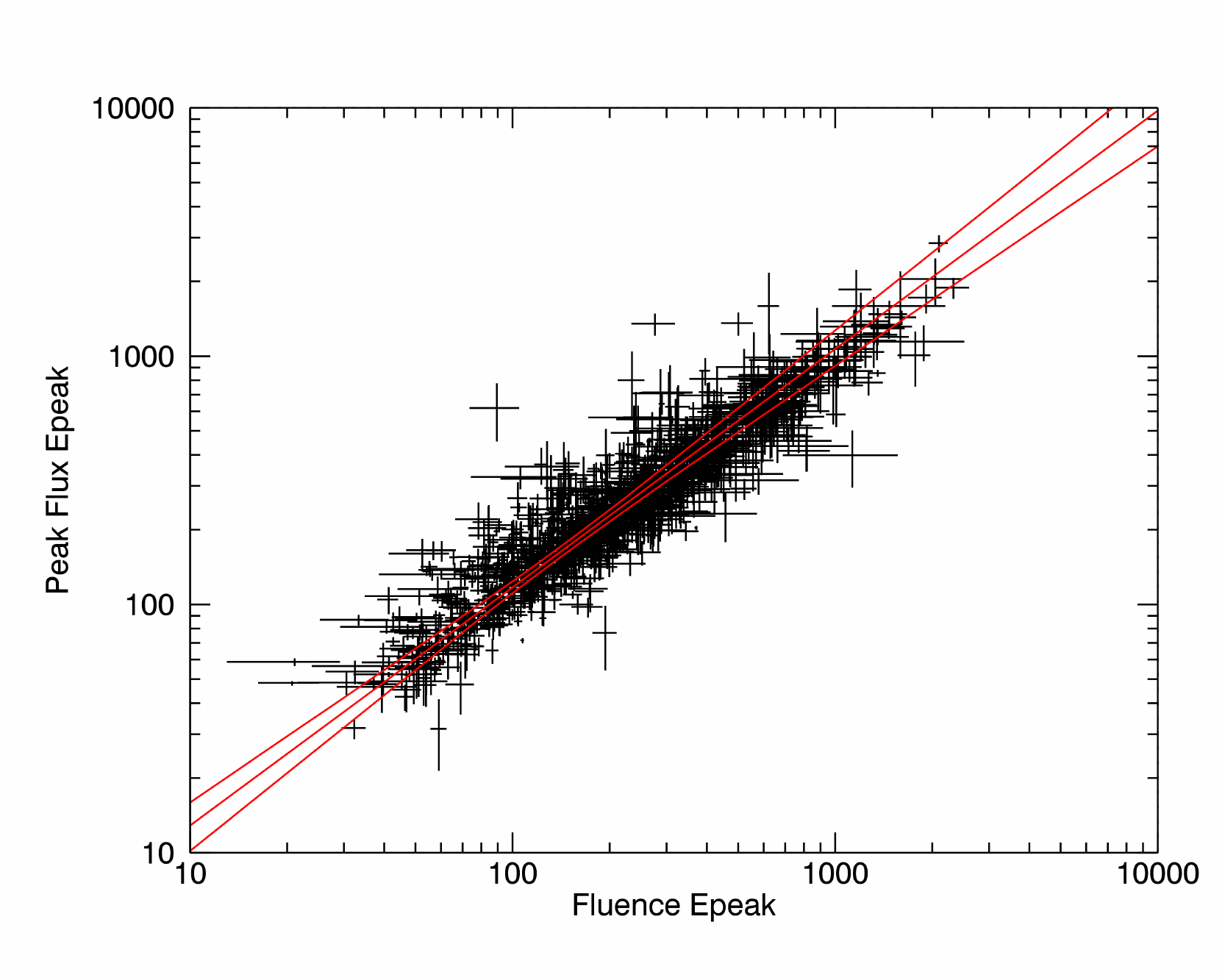}
\caption{The values of fitted \epeak\ for time-averaged ('fluence') spectra are plotted against the 
fitted \epeak\ values from the peak flux spectrum for the same burst. The 1188 GRBs are from the BATSE 
Spectroscopy Catalog \citep{GoldsteinCat}, where the best fits were either COMP or  BAND and the 
\epeak\ values were constrained. \label{EpeakCorrelation}}
\end{figure}

In GRB pulses, at least, it appears that there is a good correlation between the values of \epeak\ 
derived from fits to the peak and fluence spectra. Not all bursts are so strongly dominated by a 
single pulse in the light curve, yet we have made the case above that the spectral characteristics 
are dominated by the sampling bias at the peak of the pulse. If bursts are typically made up of any 
number of discrete and overlapping pulses, the peaks of these pulses should still dominate their 
fluence spectra. We have tested this with fits to 1188 BATSE GRBs from the complete spectral catalog 
\citep{GoldsteinCat}, as shown in Figure \ref{EpeakCorrelation}. The peak flux and fluence values are 
clearly correlated, with the best fit power law trend shown in red, along with the error bounds of 
the fit. The fitted power law function is $E_{\rm p,p} = 1.41 \times E_{\rm p,f}^{0.96}$, which also 
indicates the trend where the peak \epeak\ is larger than the fluence \epeak.
The Pearson correlation coefficient (linearity in log-space) is $0.895 \pm 0.003$, and the Spearman 
rank coefficient (monotonicity) is $0.892 \pm 0.003$. The uncertainty assumes the Epeak errors are 
(log-)normally distributed. One possible consequence of this 
correlation is the correspondence between Amati-like relations that use the fluence-based \epeak\ 
\citep{Amati02} and the Yonetoku-like relations that use the peak \epeak\ \citep{Yonetoku,Ghirlanda}: 
they are basically the same value. In our picture of strong spectral evolution within pulses, the 
\epeak\ value for the fluence spectrum is dominated by the heavy weighting by the spectra close to 
peak of the pulse, as they contain the majority fraction of the counts in the fluence. 

\subsection{The Role of Pulse Asymmetry} \label{asymsection}

We have shown above that the peak spectra dominate the fluence-derived \epeak\ value for a pulse with 
strong spectral evolution. Clearly, there should be an effect on this derived value that depends upon 
where in the pulse history the peak lies: early or late. If the pulse rises quickly, the peak spectra 
will sample the earlier values of the \epeak\ evolution trend and thus should be higher on average. The 
opposite case holds for pulses that peak later relative to the overall pulse duration.
\citet{HakkilaPreece} have shown that the majority of bright pulses in their catalog are asymmetric 
and that the asymmetry is generally correlated with spectral hardness \citep{HakkilaSwift15}.

Following the sources cited above, we define the pulse asymmetry at the level of $e^{-3}$ of the peak intensity as:
\begin{equation}
\kappa = \frac{1}{\sqrt{1 + 4 \sqrt{\tau_1 / \tau_2}/3}}.
\end{equation}
This quantity ranges from $\kappa =1 $, for an early-peaking asymmetric pulse, to $\kappa =0$, 
for a symmetric pulse (the peak divides the duration in two). Note that the two pulse shape parameters 
of the Norris pulse model ($\tau_1\ \&\ \tau_2$) do not directly map into the rise time or decay time of the pulse. 
Formally, the rise and decay times may be determined from the shape constants by the expressions
\begin{eqnarray}
\tau_{\rm rise} = & \frac{n\tau_2}{2}\left[\sqrt{1 + 4 \sqrt{\tau_1 / \tau_2}/n}- 1\right],\nonumber \\
{\rm and}\quad \tau_{\rm decay} = & \frac{n\tau_2}{2}\left[\sqrt{1 + 4 \sqrt{\tau_1 / \tau_2}/n}+ 1\right],\label{risedecay}
\end{eqnarray}
where, as above, we take $n=3$.
Indeed, the shape parameters from the last line in the Table, $\tau_1 = 150$ s and $\tau_2 = 1$ s, produce an approximately 
symmetric pulse, while the values from the first line, $\tau_1 = 6$ s and $\tau_2 = 5$ s, result in a very fast rise,
slower decay pulse with high early-peaking asymmetry. 
Pulses with late-peaking asymmetry can not be modeled (because of the constraint $\kappa \ge 0$). Late-peaking pulses are 
rare in the published literature of GRB pulse modeling.

We constructed five separate sets of synthetic GRB spectral histories, each with different values for $\tau_1$ and $\tau_2$,
resulting in five different values for $\kappa$. In each set, we created 10 bursts with different peak 
count rates: 10,000, 9,000, 8,000, 7,000, 6,000, 5,000, 4,000, 3,000, 2,000 and 500 count/s. Averaging 
the spectral fits over each set should smooth out difficulties that arise with the pulse model: for 
instance, the pulse width ($t_{\rm dur}$) is defined with respect to the 3 {\em e}-folding decrease on either side 
of the pulse peak. Depending on the overall intensity of the pulse, this value may differ significantly 
from the duration, either derived formally (e.g.: as ${\rm T}_{90}$, the duration of $90\%$ of the flux 
\citep{Koshut}) or determined by eye. When the exponential 
rise begins and the tail ends depends upon the signal to noise of each spectrum in the sequence. 

\begin{table*}
\begin{center}
\begin{tabular}{l c c c c c c c c}
\tableline\tableline
$\tau_1$ & $\tau_2$ & $\kappa$ & $\tau_{\rm peak}$ & $\tau_{\rm dur}$ & $\tau_{\rm rise}$ & $\tau_{\rm decay}$ & 10k \epeak\ & Ave.\ \epeak\ \\
(s)      & (s)      &          & (s)               & (s)                & (s)               & (s)                & (keV)       & (keV)     \\
\tableline
6 & 5 & 0.64 & 5.5 & 23.5 & 4.3 & 19.3 & $297 \pm 9$ & $303 \pm 29$  \\
3 & 5 & 0.70 & 3.9 & 21.4 & 3.2 & 18.2 & $241 \pm 5$ & $262 \pm 54$  \\
6 & 3 & 0.59 & 4.2 & 15.3 & 3.1 & 12.1 & $200 \pm 5$ & $224 \pm 43$  \\
20 & 3 & 0.47 & 7.7 & 19.0 & 5.0 & 14.0 & $153 \pm 3$ & $155 \pm 9$  \\
150 & 1 & 0.24 & 12.2 & 12.5 & 4.7 & 7.7 & $79.5 \pm 1$ & $83 \pm 6$ \\
\tableline
\end{tabular}
%\\[3.5pt]
\caption{Fluence \epeak\ and Pulse Asymmetry. A series of simulations were performed to demonstrate the effect of pulse asymmetry for a standard 
spectral evolution model during each pulse. Columns 1 and 2 contain the pulse rise and decay parameters 
($\tau_1$ \& $\tau_2$); ten simulations were created for each of these pulse shape values, where the pulse 
peaks were scaled as described in the text. The derived asymmetry ($\kappa$), the peak time ($\tau_{\rm peak}$), duration ($\tau_{\rm dur}$) and the rise and decay times 
of the pulse ($\tau_{\rm rise}$ and $\tau_{\rm decay}$) follow in columns 3 -- 7. Column 8 gives the \epeak\ value for the 
single 10,000 count/s run (the brightest in each set of ten), while the last column gives the average 
of the fitted \epeak\ values for the ten runs in each set.\label{asymmetry}}
\end{center}
\end{table*}

The results of this analysis, as presented in Table \ref{asymmetry}, show that there is a trend toward 
higher \epeak\ values for higher values of asymmetry ($\kappa$ in the Table). This trend is seen both 
for the highest count-rate simulation (`10k \epeak' column), as well as the for the average over the 
ensemble of weak to bright pulses. The relation between the pulse parameters, especially the position 
(in time) of the peak, and the derived duration, are complicated by the definition of pulse duration in 
terms of number of e-foldings of the rise and decay exponentials. As seen in the last row of the table, 
the peak time seems to be equal to the duration; in such cases, the start time of the duration is delayed. 
Accordingly, we include the rise and decay times ($\tau_{\rm rise}$ and $\tau_{\rm decay}$), as 
calculated by Equation \ref{risedecay}, in columns 6 and 7.
Still, the general trend may be described as correlation between asymmetry and \epeak: the later the 
pulse peaks, the lower the fluence-averaged \epeak\ value that will be obtained for the same power-law spectral 
evolution in a pulse.

\section{More Complicated Lightcurves}

Most GRB lightcurves are not as simple as our single-pulse model. Some consist of well-separated 
(or, at least: clearly overlapping) single pulses, such that the pulses can be individually fitted 
(see, for example, the pulse catalogs in \citet{Ford,HakkilaPreece,lu10}). A main result of  
these efforts was that individual pulses within bursts could be categorized into two broad classes 
based upon their spectral evolution: hard-to-soft (HTS) and hardness-intensity tracking (HIT), as 
discussed above. 
%The HIT behavior is manifested as a correlation between \epeak\ and some measure of 
%the intensity of each spectrum within at pulse. 
So far, we have focused on HTS pulses. As long as separable pulses can be identified within a complex light 
curve, the results we have identified so far should hold for more complex bursts, as long as each pulse 
belongs in the HTS category. The existence of the HIT category raises several issues, then: can we 
determine the relationship between the peak flux and fluence values for \epeak; and, can we account 
for HIT evolution within the context of HST evolution pulses?

As Figure \ref{sim40k_phot_flux_histo} demonstrates, burst lightcurves are the most heavily sampled at 
the peak. The result is that the time-averaged (`fluence') spectrum is heavily weighted by those spectra, 
resulting in an average fitted \epeak\ value that represents them. This should be true of pulses 
with HIT spectral evolution as well: the softer spectra at the rise and tail of the pulse are underrepresented 
in the average, just as they are in HTS pulses. In this case, instead of very hard spectra being under-sampled 
during the rising portion of the pulse, the rising portion of the HIT spectral evolution mirrors 
the decay in time, so no spectra are harder than those at the peak of the pulse. Thus, regardless of 
whether the pulse is HTS or HIT, the fluence \epeak\ should correlate well with that of the peak flux 
spectrum within a burst. HIT pulses should not destroy the correlation found in Figure \ref{EpeakCorrelation}.

\begin{figure}[t]
\includegraphics[scale=.45]{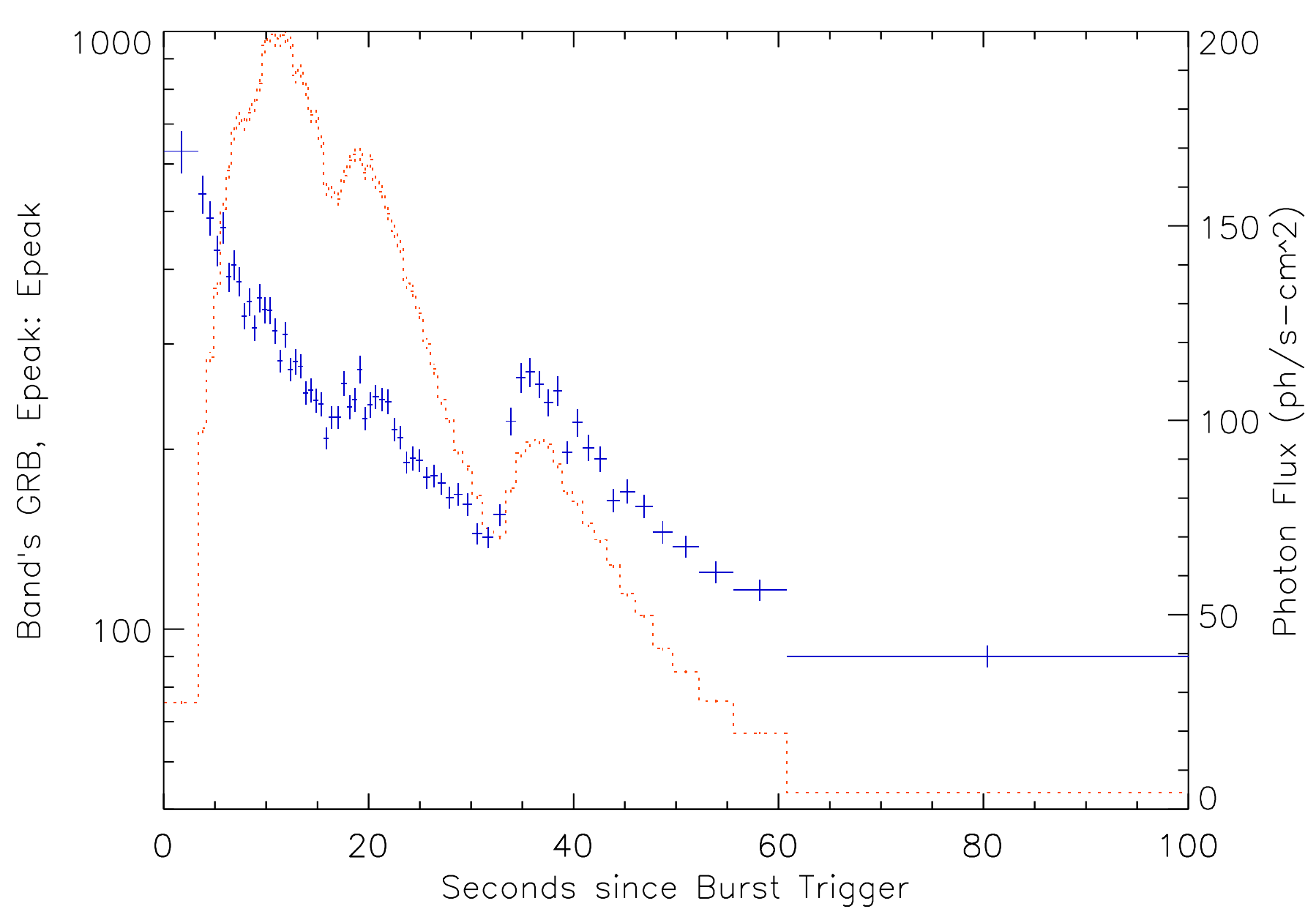}
\caption{The spectral evolution of one simulation of a complex lightcurve shows both hard-to-soft and 
hardness-intensity-tracking behavior in \epeak. The left and right axes correspond to \epeak\ (in keV) 
and photon flux, respectively. The parameters of the simulation are detailed in the text, but 
can be described as seven identical pulses with an overlap in time that is spaced in a geometric 
progression series with a common ratio of 2. \label{spec_mdl_BandsGRBEpeak_100}}
\end{figure}

More interesting is the question of generating HIT pulses within the HTS pulse paradigm. 
%This possibility was suggested by \citet{HakkilaPreece}. 
Where two HTS pulses overlap, the combined spectra are likely to be 
dominated by the more intense of the two, especially when signal-to-noise binning is taken into account. 
Thus, we should expect that the least intense, very hard spectra at the rise portion of a later pulse to be completely 
dominated by softer spectra on the decay tail of the preceding pulse. This will continue until the 
spectra from the following pulse start to dominate the spectral fits, which should occur near the peak 
intensity of the later pulse, if it rises sufficiently rapidly. 
In order to determine what happens in the transition 
region where the intensities are roughly equal but the individual pulses contribute \epeak\ values that 
are considerably different, we turn back to simulations. Figure \ref{spec_mdl_BandsGRBEpeak_100} shows 
the result of several overlapping pulses, each with the the same rise and decay times (3.0 and 11.0 s), 
the same intensities (10000 count/s), the same \epeak\ evolution parameters (1500. to 100. keV), but 
increasing start times: 0, 1, 2, 4, 8, 16, and 32 s. There is no physical motivation for this time series. 
% External shock theory predicts that later pulses should be systematically delayed; however, the later 
% pulses should also be stretched in time and diminished in intensity and energy in the same manner. 
The first four pulses blend together into one relatively intense pulse, which has not lost the overall HTS 
spectral evolution of its constituents. The last two pulses however, demonstrate classic HIT behavior. 
The possibility that isolated pulses may actually be comprised of a train of subpulses was raised by 
\citet{hakkilapreece14}, who discovered a conserved pattern of residuals after subtracting the Norris 
pulse model fit from observed pulses and stacking the residuals. There is an indication for a hardening 
in each of the sub-pulses, as observed in Figure \ref{spec_mdl_BandsGRBEpeak_100}.
A full treatment of this question is somewhat outside of the main scope of the present work; we hope to 
address it more completely in a later paper.

\section{Discussion}

Out of the many \epeak\ values that can be found in any given burst, the question of which \epeak\ can 
represent the burst to the extent that it can be an accurate cosmological indicator % seems to ultimately depend upon chance. 
has at least two robust answers: both the time-averaged and peak flux-derived values, due to their correlation.
The details of spectral evolution, pulse asymmetry and pulse shape apparently 
conspire with signal-to-noise temporal binning to pick out a medium energy value from a span that can 
cover at least a decade. 

We have constructed a simulation that reproduces the time history and spectral evolution found in 
detail for at least two bright pulses in observed GRBs, as well as qualitatively in several other 
observed single-pulse bursts. Using this pulse simulation, we are also able to determine why the 
fitted values of \epeak\ are so highly correlated between the time averaged and peak flux spectra in
GRBs: a signal-to-noise temporal binning of spectra within a burst profile strongly biases the number 
of spectra at the peak of pulses. This has the effect of creating a large number of spectra during a 
relatively narrow time window, where \epeak\ doesn't change much and is mid-way through its evolution 
from hard to soft. The peak flux spectrum picks out an \epeak\ value that is far from the extremes 
of its evolution, while the fluence spectrum has the spectral characteristics near the peak of the pulse 
as the overwhelming contributor to the average. A more subtle consequence 
of the predominance of peak spectra sampling comes in the interpretation of the \epeak\ distributions 
as presented in time-resolved spectral catalogs of GRBs (e.g. \citet{KanekoCat}). GRB spectra clearly 
can have very high and very low \epeak\ values and we have shown that the fitting process does not 
add very much dispersion to the distribution. So, how does the observed distribution with a FWHM of 
approximately a decade in energy arise? Our pulse simulations have shown that: either the limiting energy parameter 
priors are such that there are no constraints, in which case some unknown mechanism is at work to create 
the shape of observed distribution. Or else pulses act like our impulsive model, defined by the 
highest energy at the pulse leading edge, which does have the effect of selecting out the median 
energies for peak flux and fluence spectra, resulting in the observed narrow distributions.
% the numerical sampling bias toward the median values manifests 
% itself as a quite narrow distribution, with approximately a decade in energy FWHM, as seen in Figures 
% \ref{BATSE_GBM_Epeak} and \ref{sim_ep_histo}.

It is reasonable to ask whether the use of a temporal binning scheme other than signal-to-noise might have 
an impact on our results. The Bayesian Block (BB) histogram representation of time series data is a method 
that can optimally capture the statistically important features of the data \citep{scargle12}. For example, 
a Poisson-dominated time series that is constant within a predefined measure of probability would be represented by a single bin 
in the Bayesian Block scheme. This has the advantage of capturing quiescent periods between pulses in 
a single time bin that can be omitted from further analysis. For the same time series, a signal-to-noise binning might extend an  
incomplete time bin at the trailing edge of one pulse through a quiescent period into the rising 
portion of the next pulse before sufficient counts above background have been accumulated to fill out the bin. 
Thus, one scheme creates a good representation of the peaks and valleys of a complex light curve (BB), 
while the other accumulates bins of equal statistical weights (SNR). For time-resolved spectroscopy, 
the SNR representation gives binning of equal statistics, so that the spectra can be compared within 
single bursts and between different bursts. 
The BB binning would tend to assign the peak into a single bin, which is approximately constant, while 
breaking up a fast rising (or falling) portion into many bins, some of which may be sub-optimal for 
spectral fitting. Of course, the BB binning criteria can be adjusted until every bin has at least the minimum 
SNR for spectral fitting, perhaps at the expense of allowing some bins to have much higher weighting. 
So far, no comprehensive time-resolved spectroscopy catalog has used this method, although it was 
a feature of the analysis of a selection of single-pulse bursts by \citet{burgess14}. This approach 
allowed for excellent determination of the spectral parameters of their model.

The HTS spectral evolution we have considered here as an archetype for the GRB pulse profile has quite 
profound physical implications. In particular, the higher energies are concentrated (by the dual action of narrowing pulse widths 
and decreasing spectral lags) into shorter intervals of time occurring earlier in the pulse. This is 
borne out in the case of extremely high energies in the first pulse of GRB 130427A, where three LAT photons 
($> 100$ MeV) are observed within 0.1 s either way around the GBM trigger time, followed by a gap of roughly 5 s., 
during which only one other photon above 100 MeV is observed \citep{preece2014}. There is a wider 
distribution of LAT LLE photons (30 -- 100 MeV), with a FWHM of roughly 0.4 s surrounding the higher energy 
LAT photons \citep{Ackermann}. Clearly, bursts are {\em capable} of producing the highest energy photons impulsively. 
That being the case, questions arise concerning the nature of the pulse as seen in lower energies,
as covered by BATSE or GRB NaI detectors. Perhaps the entire evolution of the pulse driven by an impulsive 
event that somehow rapidly fills up a reservoir of energy that is released over time, as proposed by 
\citet{LiangKargatis96}, \citet{KocevskiLiang} and \citet{BasakRao12}. In accounting for the spectral 
lag evolution in GRB pulses as a function of the running integrated photon fluence,
\begin{equation}
E_{\rm peak} = E_0e^{-\Phi(t)/\Phi_0},
\label{epfluence}
\end{equation}
these authors do not address the {\em pulse} flux evolution that drives the running fluence $\Phi(t)$. 
Left unanswered is why there should be a rising edge to the pulse, rather than, say, an abrupt turn-on 
or even a simple decaying exponential shape, neither of which is commonly observed in GRBs, if at all. 
In our impulsive model, the pulse rise occurs as the bulk of the cooling photons enter into the detector 
passband from above, until the peak, when all the higher energy photons have been degraded. The rest of 
the lightcurve is then seen as a cooling curve, with some photons moving into the X-ray band and dropping 
out of view. A impulsively illuminated, thin-shell, relativistic flow exhibits rise, 
decay and asymmetric pulse behavior simply due to curvature effects, as described by \citet{Dermer2004}. 
In addition, a kinematic treatment predicts that the instantaneous $\nu\cal{F}_{\nu}$ flux should behave as 
$E_{\rm peak}^3$ during the pulse decay phase. This seems to be at odds with some of the observed pulse spectral 
evolution \citep{BorgonovoRyde}. \citet{preece2014} have addressed this by invoking magnetic reconnection 
or mini-jets as the impulsive event, followed by magnetic flux freezing in the expanding fluid element.

\begin{figure}[t]
\includegraphics[scale=.45]{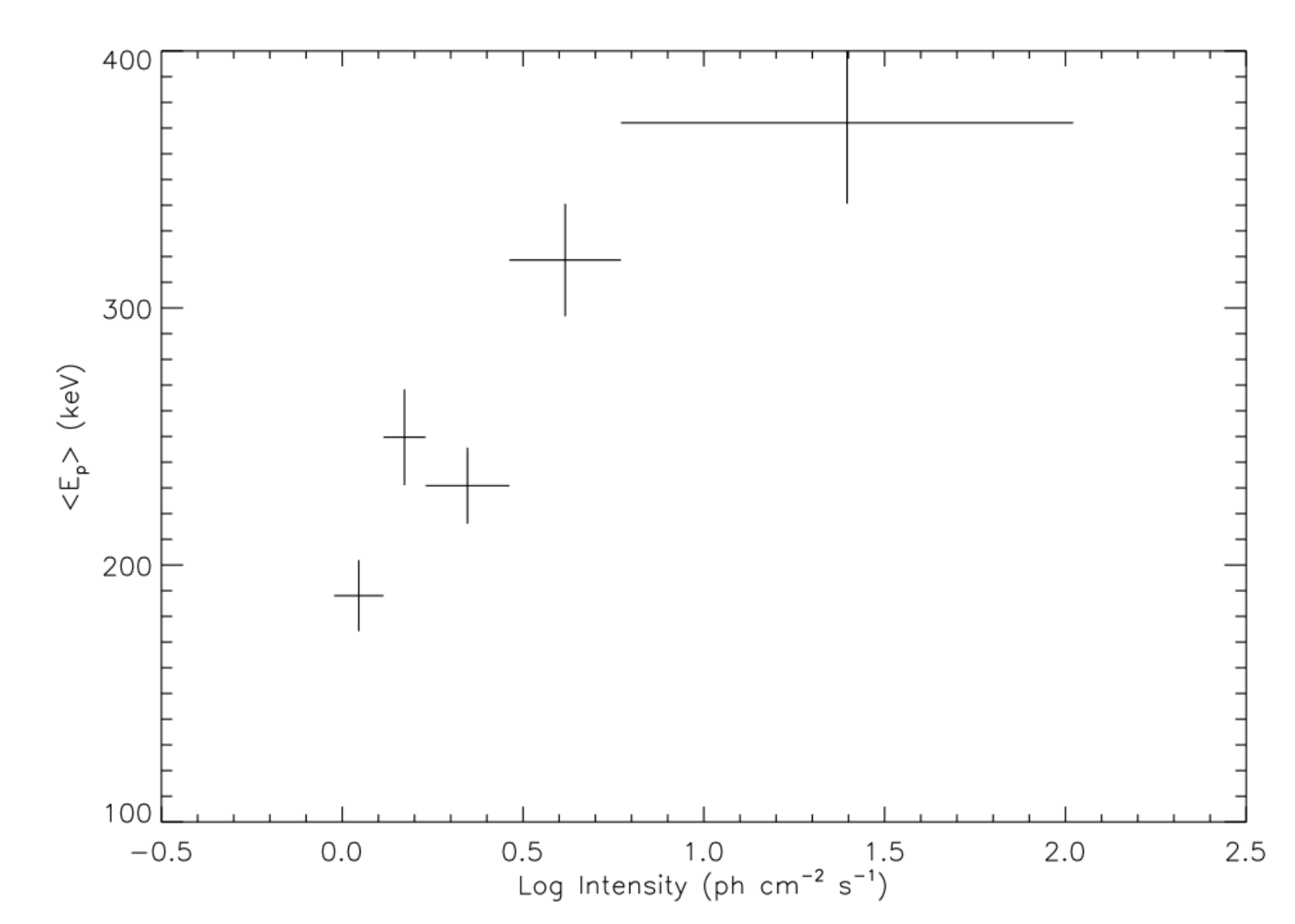}
\caption{The average peak $\nu{\cal F}_{\nu}$ energies as a function of intensity for five groups of 
peak flux spectra from 1421 BATSE GRBs. The horizontal error bars indicate the intensity bin widths, while the vertical 
error bars represent the the $1\sigma$ errors in the mean, where the \epeak\ distributions were assumed 
to be approximately Gaussian in log energy. \label{Ep-PeakFlux3}}
\end{figure}

The less intense a pulse is (as in dimmer GRBs), the 
more bins around the pulse peak must be included to maintain a constant SNR for spectroscopy; however, 
given the nearly constant spectral evolution at the peak, the larger number of bins included in the fit 
does not affect the mean \epeak. With HTS spectral evolution, we expect the peak \epeak\ to be robust 
with respect to intensity for even the broadest binning required for the dimmest bursts. 
We have done an analysis of the peak flux spectra from the 50 simulated pulses described in Table 
\ref{asymmetry} (which divides into ten peak flux groups) that shows no change in average 
\epeak\ values from dimmest to brightest: the slope in average \epeak\ as a function of peak flux is only $0.1 \sigma$ 
different from 0, with $99.99\%$ confidence. Thus, in the model where bursts are composed entirely from 
separable pulses (even if they can not be separated formally), observed correlations between peak 
flux \epeak\ and peak intensity are almost certainly intrinsic in nature. 
With 399 bursts from an early version of the BATSE Spectroscopy Catalog, \citet{Mallozzi} demonstrated 
that there is a significant correlation between the peak intensity and mean \epeak\ for five intensity 
groups. We have repeated this analysis using the complete catalog with 1421 bursts 
(\citet{GoldsteinCat}, or 3.5 times as many bursts as in the Mallozzi data set), 
as shown in Figure \ref{Ep-PeakFlux3}. The strong shift in mean \epeak\ 
as a function of intensity that is replicated here must be intrinsic, and follows the trend expected 
from standard cosmology, as was pointed out by \citet{Mallozzi}. 

\section{Conclusions}

Given that pulse peaks of GRBs pick out a representative value of \epeak, the answer to the question posed in the Abstract, 
``Which \epeak?'' is: ``The peak flux \epeak''. We have shown that a simple pulse model with HTS spectral 
evolution can account for several observables: 
\begin{itemize}
\item the narrow \epeak\ distributions found in several spectral catalogs;
\item the fluence and peak flux \epeak\ correlation;
\item the role of pulse asymmetry on the average \epeak;
\item the constraint of the temporal decay index to observed values for widely different values of $E_{\rm hi}$;
\item the impulsive nature of pulse energization; and
\item the frequent observation that the first spectrum in a burst is the hardest.
\end{itemize}
In addition, these results are robust, even to the point that we need only assume that 
a pulse has {\em some} type of spectral evolution, either HTS or HIT. The latter may be gotten by layering 
HTS pulses on top of each other. Finally, bursts with multiple pulses can be thought of a collection of 
average spectra for the individual pulses, each of which is represented by the average \epeak\ at the peak. 
We don't yet have
a theory for the spectral evolution of {\em pulses}; however, to the extent that bursts are generally
HTS, the peak \epeak\ of this distribution, as observed in the spectral window of the detector,
should dominate the burst itself.

\section*{Acknowledgements}
We are grateful for the comments on the first draft of this manuscript made by the anonymous referee. 
We had overlooked a crucial aspect of the analysis and the paper is much improved as a result.

%%References

\end{document}